\documentclass[aps,prd,preprint,superscriptaddress,showpacs]{revtex4}
\usepackage{graphicx}

\def\bwt{\begin{widetext}}
\def\ewt{\end{widetext}}
\def\be{\begin{equation}}
\def\ee{\end{equation}}
\def\bea{\begin{eqnarray}}
\def\eea{\end{eqnarray}}
\def\bean{\begin{eqnarray*}}
\def\eean{\end{eqnarray*}}
\def\bary{\begin{array}}
\def\eary{\end{array}}
\def\bit{\begin{itemize}}
\def\eit{\end{itemize}}

\def\su5u1{SU(5) \times U(1)}
\def\fsu5u1{SU(5) \times U(1)'}
\def\so10{SO(10)}
\def\sq20{SO(10) \times SO(10)}

\usepackage[centertags]{amsmath}
\usepackage{amssymb}
\newcommand{\Z}{{\mathbb Z}}

\begin{document}

\title{Type IIA Pati-Salam Flux Vacua}

\author{Ching-Ming Chen}

\affiliation{George P. and Cynthia W. Mitchell Institute for
Fundamental Physics,
 Texas A$\&$M University, College Station, TX 77843, USA }

\author{Tianjun Li}

\affiliation{Department of Physics and Astronomy, Rutgers University,
Piscataway, NJ 08854, USA }

\affiliation{Institute of Theoretical Physics, Chinese Academy of Sciences,
 Beijing 100080, P. R. China }

 \author{Dimitri V. Nanopoulos}

\affiliation{George P. and Cynthia W. Mitchell Institute for
Fundamental Physics,
 Texas A$\&$M University, College Station, TX 77843, USA }

\affiliation{Astroparticle Physics Group,
Houston Advanced Research Center (HARC),
Mitchell Campus, Woodlands, TX 77381, USA}

\affiliation{Academy of Athens, Division of Natural Sciences,
 28 Panepistimiou Avenue, Athens 10679, Greece }

\date{\today}

\begin{abstract}

We show that for supersymmetric AdS vacua on Type IIA orientifolds
with flux compactifications, the RR tadpole cancellation conditions
can be completely relaxed, and then the
four-dimensional $N=1$ supersymmetry conditions are the main
constraints on consistent intersecting D6-brane model building.
We construct two kinds of three-family Pati-Salam models.
In the first kind of models, the suitable three-family
SM fermion masses and mixings can be generated at the stringy
tree level, and then the rank one problem for the SM
fermion Yukawa matrices can be solved. In the second kind
of models, only the third family of the SM fermions
can obtain masses at tree level. In these models, the
complex structure parameters can be determined by supersymmetric
D6-brane configurations, and all the moduli may be stabilized.
The initial gauge symmetries $U(4)_C \times U(2)_L \times U(2)_R$ and
$U(4)_C \times USp(2)_L \times U(2)_R$ can be
broken down to the $SU(3)_C\times SU(2)_L\times U(1)_{B-L} \times
U(1)_{I_{3R}}$ due to  the Green-Schwarz mechanism
and D6-brane splittings, and further down to the
SM gauge symmetry around the string scale via the supersymmetry
preserving Higgs mechanism.
Comparing to the previous model building, we have less bidoublet
Higgs fields. However, there generically exist some exotic particles.

\end{abstract}


\preprint{ACT-01-06, MIFP-06-02, hep-th/0601064}

\maketitle

\section{Introduction}
The constructions of realistic Standard-like string models
with moduli stabilization
is the major challenge and lasting problem in string phenomenology.
In the beginning
string model building was mainly concentrated on the weakly
coupled heterotic string theory, and rather successful
models like flipped $SU(5)$~\cite{Barr:1981qv} in its stringy form
were constructed~\cite{AEHN}. After the
second string revolution we can construct
consistent four-dimensional chiral models with
non-Abelian gauge symmetry on Type II orientifolds
because of the advent of  D-branes~\cite{JPEW}.

During last a few years, Type II orientifolds with intersecting D-branes
has been particular interesting in string model building where
the chiral fermions arise from the intersections of D-branes
in the internal space~\cite{bdl} with T-dual description in terms
of magnetized D-branes~\cite{bachas}.
On Type IIA orientifolds with intersecting D6-branes,
 a large number of non-supersymmetric three-family
Standard-like models and grand unified models, which
satisfy the Ramond-Ramond (RR) tadpole
cancellation conditions, were
constructed~\cite{Blumenhagen:2000wh,Angelantonj:2000hi,Blumenhagen:2005mu}.
However, there generically exist the
uncancelled Neveu-Schwarz-Neveu-Schwarz (NSNS) tadpoles
and the gauge hierarchy problem. To solve these two problems,
the first quasi-realistic supersymmetric models~\cite{CSU1,CSU2}
 have been constructed in Type IIA theory on
$\mathbf{T^6/(\Z_2\times \Z_2)}$ orientifold with intersecting D6-branes.
Then, the supersymmetric Standard-like models,
 Pati-Salam models, $SU(5)$ models as well as
 flipped $SU(5)$ models have been constructed
systematically~\cite{CP,Cvetic:2002pj,CLL,Cvetic:2004nk,Chen:2005ab,Chen:2005mj}, and
their  phenomenological consequences have been
studied~\cite{CLS1,CLW}.
Moreover, the supersymmetric constructions on other
orientifolds were also discussed~\cite{ListSUSYOthers}.
There are two main constraints on supersymmetric model
building: RR tadpole
cancellation conditions  and four-dimensional $N=1$ supersymmetry conditions.

Even though some of the
complex structure parameters (in  the Type IIA picture)
and dilaton field may be stabilized due to gaugino
condensation in the hidden sector in some models
(for example, see~\cite{CLW}),
the moduli stabilization in open string and
closed string sectors is still one of the most
difficult problems.
With supergravity fluxes which introduce a supergravity
potential, we can stabilize the
compactification moduli
fields by lifting the continuous moduli space of the string vacua in
 the four-dimensional effective theory
(for example, see~\cite{GVW}).
On Type IIB orientifolds, the supergravity fluxes
contribute large positive D3-brane charges
due to the Dirac quantization conditions. Then, they
 modify the global RR tadpole cancellation conditions
significantly and imposes strong constraints on
consistent model building~\cite{CU,BLT}.
By introducing magnetized D9-branes with
large negative D3-brane charges in the hidden
and observable sectors, we can construct the three-family and four-family
Standard-like
models~\cite{Chen:2005mj,MS,CL,Cvetic:2005bn,Kumar:2005hf,Chen:2005cf}.

However, in above model building with or without fluxes,
it is very difficult to explain the three-family
Standard Model (SM) fermion masses and
mixings. In the $SU(5)$ models,  flipped $SU(5)$ models,
and trinification models, the up-type
quark Yukawa couplings, down-type quark Yukawa couplings,
and lepton Yukawa couplings are
forbidden by the anomalous $U(1)$ symmetries, respectively.
And for the Pati-Salam models, although all the SM fermion Yukawa couplings
can in principle be allowed by the anomalous $U(1)$ symmetries,
we may not give masses to the three families of the SM fermions
at the stringy ($M_S\sim 10^{17}$ GeV) tree level. 
The point is that the left-handed SM fermions,
 right-handed SM fermions and bidoublet Higgs fields in general arise from
the intersections on different two-tori, and
then the SM fermion Yukawa matrices are generically rank one,
{\it i.~e.}, only the
third family of the SM fermions can obtain masses at the stringy
tree level~\cite{CLLL-P}.
This is so called ``rank one problem'' in intersecting D-brane
model building (for a possible solution, see~\cite{Dutta:2005bb}).
In addition, if  supersymmetry is broken by supergravity fluxes,
  the massless SM fermions
may not obtain masses from quantum corrections~\cite{CLLL-P}
 because the supersymmetry breaking trilinear soft terms are
universal and the supersymmetry breaking soft masses for the left/right-chiral
squarks and sleptons are
universal~\cite{Soft BreakingI,Soft BreakingII,Soft BreakingIII}
since the K\"ahler potential for
the SM fermions depends only on the intersection
angles~\cite{CPY,LustetalIII,Abel:2004ue,Bertolini:2005qh}.
Thus, how to construct the 
Standard-like models, which can generate the suitable three-family SM fermion
masses and mixings at the stringy tree level, is a very interesting open question.

Recently, the techniques for consistent chiral flux compactifications on
Type IIA orientifolds with intersecting D6-branes were
developed~\cite{Grimm:2004ua,Villadoro:2005cu,Camara:2005dc,Camara:2005pr}.
 We will show that in supersymmetric AdS vacua,
the metric, NSNS and RR fluxes can contribute
negative D6-brane charges to all the RR tadpole cancellation conditions,
{\it i.~e.}, the RR tadpole cancellation conditions give no constraints
on consistent model building. Thus, the supersymmetric flux models
on Type IIA orientifolds are mainly constrained by four-dimensional
$N=1$ supersymmetry conditions, and then
 it is possible to construct the rather realistic 
intersecting D6-brane models.

In this paper, we construct two kinds of three-family Pati-Salam models
on Type IIA orientifolds with flux compactifications in supersymmetric
AdS vacua. In the first kind of models, three families of the
SM fermions can obtain suitable masses at the stringy tree level,
while in the second kind of models,
the third family of the SM fermions can obtain
masses at tree level, and we assume that the masses for the
first two families of the SM fermions
may be generated from quantum corrections. In these models,
we can determine  the complex structure parameters via supersymmetric
D6-brane configurations, and may stabilize all the moduli.
The initial gauge symmetries are $U(4)_C \times U(2)_L \times U(2)_R$ and
$U(4)_C \times USp(2)_L \times U(2)_R$ (note that $USp(2)_L\equiv SU(2)_L$).
These initial gauge symmetries can be
broken down to the $SU(3)_C\times SU(2)_L\times U(1)_{B-L} \times
U(1)_{I_{3R}}$ due to  the generalized Green-Schwarz mechanism
and  D6-brane splittings, and further down to the
SM gauge symmetry at about the string scale via the supersymmetry
preserving Higgs mechanism,  where the
Higgs fields come from the massless open string states in a $N=2$
subsector. Moreover, if we want to use supergravity fluxes to relax the RR
tadpole cancellation conditions in model building
in Type IIA theory on $\mathbf{T^6/(\Z_2\times \Z_2)}$ orientifold,
the fluxes will contribute large negative D6-brane charges
due to the Dirac quantization conditions. Then
we have to introduce many D6-branes in the hidden sector
to completely cancel the RR tadpoles, and thus may introduce
a lot of exotic particles simultaneously. Therefore, we mainly
consider the Pati-Salam models in Type IIA theory on
$\mathbf{T^6}$ orientifold with flux compactifications.
We emphasize that all our Pati-Salam models on
Type IIA $\mathbf{T^6}$ orientifold
can be similarly constructed on Type IIA
 $\mathbf{T^6/(\Z_2\times \Z_2)}$ orientifold by introducing more
stacks of D6-branes in the hidden sector.

For the first kind of Pati-Salam models with $U(4)_C\times U(2)_L\times U(2)_R$
gauge symmetry, we present five models (Models TI-U-i with
i=1, ..., 5) on Type IIA
$\mathbf{T^6}$ orientifold and two models (Models TI-U-6 and TI-U-7) on
Type IIA $\mathbf{T^6/(\Z_2\times \Z_2)}$ orientifold.
Also, there are   six bidoublet Higgs fields
in Models TI-U-i with
i=1, ..., 4. There are twelve pairs of vector-like bidoublet Higgs fields
from the massless open string states in a $N=2$ subsector
in  Model TI-U-5, and six pairs of vector-like bidoublet Higgs fields
in Models TI-U-6 and TI-U-7. In particular, Model TI-U-6
is the only model in this paper that
the supergravity  fluxes do not contribute D6-brane
charges, and then do not affect the RR tadpole cancellation
conditions. Because Model  TI-U-7 is on
Type IIA $\mathbf{T^6/(\Z_2\times \Z_2)}$ orientifold,
 its supergravity  fluxes do
contribute large negative D6-brane RR tadpoles due to the
Dirac quantization conditions, and then there
are many exotic particles
from extra gauge  groups due to the RR tadpole cancellation
conditions. In addition, for the first kind of Pati-Salam models with
$U(4)_C\times USp(2)_L\times U(2)_R$ gauge symmetry, we
give four models (TI-Sp-j with j=1, ..., 4) on Type IIA
$\mathbf{T^6}$ orientifold. We have three bidoublet Higgs fields
in  Models TI-Sp-1, TI-Sp-2 and TI-Sp-3, and three pairs of
vector-like bidoublet Higgs fields  in Model TI-Sp-4.

For the second kind of Pati-Salam models with
 $U(4)_C\times U(2)_L\times U(2)_R$
gauge symmetry, we present six models (Models TII-U-i with
i=1, ..., 6) on Type IIA
$\mathbf{T^6}$ orientifold. There are  two bidoublet Higgs fields
in Model TII-U-1, three in
Model  TII-U-2, four in Models TII-U-3 and TII-U-4.
And there are two and four pairs of
vector-like bidoublet Higgs fields
in Models TII-U-5 and TII-U-6, respectively.
For the second kind of Pati-Salam models with
$U(4)_C\times USp(2)_L\times U(2)_R$ gauge symmetry,
we give four models (Models TII-Sp-i with i=1, ..., 4) on Type IIA
$\mathbf{T^6}$ orientifold, and one model (Model TII-Sp-5)
on Type IIA $\mathbf{T^6/(\Z_2\times \Z_2)}$
orientifold. Also, we have one bidoublet Higgs field
in  Model TII-Sp-1, three in Models
 TII-Sp-2, TII-Sp-3, and TII-Sp-5, and four pairs of
vector-like bidoublet Higgs fields
in Model TII-Sp-4. Similar to the Model TI-U-7,
we have quite a few exotic particles in Model TII-Sp-5.
Comparing to the previous Pati-Salam model building
in Type IIA theory on $\mathbf{T^6/(\Z_2\times \Z_2)}$
orientifold without supergravity fluxes, we
have less bidoublet Higgs fields.

Furthermore, the generic feature of our models is that
there exist exotic particles. The phenomenological
consequences, for example, the SM fermion masses and
mixings, the moduli stabilization, and how to give
masses to exotic particles, will be discussed elsewhere~\cite{CLN}.

This paper is organized as follows. In Section II,
we  briefly review  the intersecting D6-brane model
building on Type IIA  orientifolds
with flux compactifications. We study the general
conditions for Pati-Salam model building in Section III. 
We discuss the first and second kinds of Pati-Salam
models in Section IV.
Discussion and conclusions are given in Section V.
In Appendices A and B, we present the
 D6-brane configurations and intersection numbers
for  the first and second kinds of Pati-Salam models,
respectively.

\section{Flux Model Building on Type IIA orientifolds}

We briefly review the rules for the
intersecting D6-brane model building in Type IIA theory on
$\mathbf{T^6/(\Z_2\times \Z_2)}$ orientifold
with flux compactifications~\cite{Villadoro:2005cu,Camara:2005dc}.
Because the model building rules in Type IIA theory on
$\mathbf{T^6}$ orientifold with flux compactifications
are quite similar, we only explain the differences for
simplicity.

\subsection{Type IIA Theory on $\mathbf{T^6/(\Z_2\times \Z_2)}$ Orientifold}

 We consider $\mathbf{T^6}$ to be a
six-torus factorized as 
$\mathbf{T^6} = \mathbf{T^2} \times \mathbf{T^2} \times \mathbf{T^2}$
whose complex coordinates are $z_i$, $i=1,\; 2,\; 3$ for the
$i$-th two-torus, respectively. The $\theta$ and $\omega$
generators for the orbifold group $\Z_{2} \times \Z_{2}$
 act on the complex coordinates as following
\begin{eqnarray}
& \theta: & (z_1,z_2,z_3) \to (-z_1,-z_2,z_3)~,~ \nonumber \\
& \omega: & (z_1,z_2,z_3) \to (z_1,-z_2,-z_3)~.~\,
\label{Z2Z2}
\end{eqnarray}
We implement an orientifold projection $\Omega R$, where $\Omega$
is the world-sheet parity, and $R$ acts on the complex coordinates as
\begin{equation}
R:(z_1,z_2,z_3)\rightarrow(\overline{z}_1,\overline{z}_2,\overline{z}_3)~.~\,
\end{equation}

Thus, we have four kinds of orientifold 6-planes (O6-planes) under
the actions of $\Omega R$,
$\Omega R\theta$, $\Omega R \omega$, and $\Omega R\theta\omega$,
respectively. In addition, we introduce some stacks of D6-branes
 which wrap on the factorized three-cycles.
There are two kinds of complex structures
consistent with orientifold projection for a two-torus --
rectangular and tilted~\cite{CSU2,LUII}. If we denote the
homology classes of the three cycles wrapped by $a$ stack of $N_a$ D6-branes
 as $n_a^i[a_i]+m_a^i[b_i]$ and $n_a^i[a'_i]+m_a^i[b_i]$
with $[a_i']=[a_i]+\frac{1}{2}[b_i]$ for the rectangular and
tilted two-tori respectively, we can label a generic one cycle by
$(n_a^i,l_a^i)$ in which
 $l_{a}^{i}\equiv m_{a}^{i}$ for a rectangular two-torus
while $l_{a}^{i}\equiv 2\tilde{m}_{a}^{i}=2m_{a}^{i}+n_{a}^{i}$ for
a tilted two-torus~\cite{Cvetic:2002pj}.
For $a$ stack of $N_a$ D6-branes along
the cycle $(n_a^i,l_a^i)$, we also need to include their $\Omega
R$ images $N_{a'}$ with wrapping numbers $(n_a^i,-l_a^i)$. For the
D6-branes on the top of O6-planes, we count them and their $\Omega R$
images independently. So, the homology three-cycles for $a$ stack
of $N_a$ D6-branes and its orientifold image $a'$ are
\begin{eqnarray}
[\Pi_a]=\prod_{i=1}^{3}\left(n_{a}^{i}[a_i]+2^{-\beta_i}l_{a}^{i}[b_i]\right),\;\;\;
\left[\Pi_{a'}\right]=\prod_{i=1}^{3}
\left(n_{a}^{i}[a_i]-2^{-\beta_i}l_{a}^{i}[b_i]\right)~,~\,
\end{eqnarray}
where $\beta_i=0$ if the $i$-th two-torus is rectangular and
$\beta_i=1$ if it is tilted. And the homology three-cycles wrapped
by the four O6-planes are
\begin{eqnarray}
\Omega R: [\Pi_{\Omega R}]= 2^3
[a_1]\times[a_2]\times[a_3]~,~\,
\end{eqnarray}
\begin{eqnarray}
 \Omega R\omega:
[\Pi_{\Omega
R\omega}]=-2^{3-\beta_2-\beta_3}[a_1]\times[b_2]\times[b_3]~,~\,
\end{eqnarray}
\begin{eqnarray}
 \Omega R\theta\omega: [\Pi_{\Omega
R\theta\omega}]=-2^{3-\beta_1-\beta_3}[b_1]\times[a_2]\times[b_3]~,~\,
\end{eqnarray}
\begin{eqnarray}
 \Omega R\theta:  [\Pi_{\Omega
R}]=-2^{3-\beta_1-\beta_2}[b_1]\times[b_2]\times[a_3]~.~\,
\label{orienticycles}
\end{eqnarray}
Therefore, the intersection numbers are
\begin{eqnarray}
I_{ab}=[\Pi_a][\Pi_b]=2^{-k}\prod_{i=1}^3(n_a^il_b^i-n_b^il_a^i)~,~\,
\end{eqnarray}
\begin{eqnarray}
I_{ab'}=[\Pi_a]\left[\Pi_{b'}\right]=-2^{-k}\prod_{i=1}^3(n_{a}^il_b^i+n_b^il_a^i)~,~\,
\end{eqnarray}
\begin{eqnarray}
I_{aa'}=[\Pi_a]\left[\Pi_{a'}\right]=-2^{3-k}\prod_{i=1}^3(n_a^il_a^i)~,~\,
\end{eqnarray}
\begin{eqnarray}
 {I_{aO6}=[\Pi_a][\Pi_{O6}]=2^{3-k}(-l_a^1l_a^2l_a^3
+l_a^1n_a^2n_a^3+n_a^1l_a^2n_a^3+n_a^1n_a^2l_a^3)}~,~\,
\label{intersections}
\end{eqnarray}
 where $[\Pi_{O6}]=[\Pi_{\Omega
R}]+[\Pi_{\Omega R\omega}]+[\Pi_{\Omega
R\theta\omega}]+[\Pi_{\Omega R\theta}]$ is the sum of O6-plane
homology three-cycles wrapped by the four O6-planes,
 and $k=\beta_1+\beta_2+\beta_3$ is
the total number of tilted two-tori.

\begin{table}[h]
\renewcommand{\arraystretch}{1.5}
\center
\begin{tabular}{|c||c|}
\hline

Sector & Representation   \\ \hline \hline

$aa$ & $U(N_a /2)$ vector multiplet and 3 adjoint chiral multiplets \\
\hline

$ab+ba$ & $I_{ab}$ $ (\frac{N_a}{2},
\frac{\overline{N_b}}{2})$ chiral multiplets  \\ \hline

$ab'+b'a$ & $I_{ab'}$ $ (\frac{N_a}{2},
\frac{N_b}{2})$ chiral multiplets  \\ \hline

$aa'+a'a$ & $  \frac 12 (I_{aa'} + \frac 12 I_{aO6}) $
anti-symmetric chiral multiplets  \\

  & $\frac 12 (I_{aa'} - \frac 12 I_{aO6})$
symmetric chiral multiplets  \\ \hline

\hline

\end{tabular}
\caption{The general spectrum for the intersecting D6-brane model building
 in Type IIA  theory on $\mathbf{T^6/(\Z_2\times \Z_2)}$ orientifold
with flux compactifications.}
\label{Spectrum}
\end{table}

For $a$ stack of $N_a$ D6-branes and its $\Omega R$ image, we have
$U(N_a/2)$ gauge symmetry, while for $a$ stack of $N_a$ D6-branes
and its $\Omega R$ image
on the top of O6-plane, we obtain $USp(N_a)$ gauge symmetry.
The general spectrum of D6-branes' intersecting at generic angles,
which is valid for both rectangular and tilted two-tori, is given
in Table \ref{Spectrum}. The four-dimensional $N=1$
supersymmetric models on Type IIA orientifolds with intersecting
D6-branes are mainly constrained in two aspects:
four-dimensional $N=1$  supersymmetry conditions, and
RR tadpole cancellation conditions.

To simplify the notation, we define the following products of wrapping
numbers
\begin{eqnarray}
\begin{array}{rrrr}
A_a \equiv -n_a^1n_a^2n_a^3, & B_a \equiv n_a^1l_a^2l_a^3,
& C_a \equiv l_a^1n_a^2l_a^3, & D_a \equiv l_a^1l_a^2n_a^3, \\
\tilde{A}_a \equiv -l_a^1l_a^2l_a^3, & \tilde{B}_a \equiv
l_a^1n_a^2n_a^3, & \tilde{C}_a \equiv n_a^1l_a^2n_a^3, &
\tilde{D}_a \equiv n_a^1n_a^2l_a^3.\,
\end{array}
\label{variables}
\end{eqnarray}

(1) {\it Four-Dimensional $N=1$  Supersymmetry Conditions}

The four-dimensional $N=1$ supersymmetry  can be preserved by the
orientation projection ($\Omega R$) if and only if the rotation angle of any
D6-brane with respect to any O6-plane is an element of
$SU(3)$~\cite{bdl}, {\it i.~e.},
$\theta_1+\theta_2+\theta_3=0 $ mod $2\pi$, where $\theta_i$ is
the angle between the $D6$-brane and the O6-plane on the
$i$-th two-torus. Then the supersymmetry conditions can be
rewritten as~\cite{Cvetic:2002pj}
\begin{eqnarray}
x_A\tilde{A}_a+x_B\tilde{B}_a+x_C\tilde{C}_a+x_D\tilde{D}_a~=~0~,~\,
\label{susyconditions}
\end{eqnarray}
\begin{eqnarray}
 A_a/x_A+B_a/x_B+C_a/x_C+D_a/x_D~<~0~,~\,
\label{susyconditionsII}
\end{eqnarray}
where $x_A=\lambda,\;
x_B=\lambda 2^{\beta_2+\beta3}/\chi_2\chi_3,\; x_C=\lambda
2^{\beta_1+\beta3}/\chi_1\chi_3,$ and $x_D=\lambda
2^{\beta_1+\beta2}/\chi_1\chi_2$ in which $\chi_i=R^2_i/R^1_i$ are the
complex structure parameters and $\lambda$ is a positive real number.

(2) {\it RR Tadpole Cancellation Conditions}

 The total RR charges from the D6-branes and O6-planes and from the metric, NSNS,
and RR  fluxes must vanish since the RR field flux lines are conserved.
With the filler branes on the top of the four O6-planes, we obtain the RR
tadpole cancellation conditions~\cite{Villadoro:2005cu,Camara:2005dc}:
\begin{eqnarray}
2^k N^{(1)} - \sum_a N_a A_a + {1\over 2}(m h_0  + q_1 a_1 + q_2 a_2 + q_3
a_3)& = & 16 ~,~\,
\label{tadodhz2z2I}
\end{eqnarray}
\begin{eqnarray}
-2^{\beta_1} N^{(2)} + \sum_a 2^{-\beta_2-\beta3} N_a B_a
+ {1\over 2} (m h_1 - q_1 b_{11} - q_2 b_{21} -
q_3 b_{31}) & = & -2^{4-\beta_2-\beta_3} ~,~\,
\label{tadodhz2z2II}
\end{eqnarray}
\begin{eqnarray}
-2^{\beta_2} N^{(3)} +\sum_a 2^{-\beta_1-\beta_3} N_a C_a
+ {1\over 2} ( m h_2 - q_1 b_{12} - q_2 b_{22} - q_3 b_{32}) & = & - 2^{4-\beta_1-\beta_3} ~,~\,
\label{tadodhz2z2III}
\end{eqnarray}
\begin{eqnarray}
-2^{\beta_3} N^{(4)}
+\sum_a 2^{-\beta_1-\beta_2} N_a D_a + {1\over 2} (m h_3 - q_1 b_{13} - q_2 b_{23} -
q_3 b_{33}) & = & - 2^{4-\beta_1-\beta_2} ~,~\,
\label{tadodhz2z2IV}
\end{eqnarray}
where $2 N^{(i)}$ are the number of filler branes wrapping along
the $i$-th O6-plane which is defined in Table \ref{orientifold}.
In addition, $a_i$ and $b_{ij}$ arise from the metric
fluxes, $h_0$ and $h_i$ arise from the NSNS fluxes, and $m$ and
$q_i$ arise from the RR fluxes. We consider these
fluxes ($a_i$, $b_{ij}$, $h_0$, $h_i$, $m$ and $q_i$)
quantized in units of 8 so that we can avoid
the problems with flux Dirac quantization conditions.

\renewcommand{\arraystretch}{1.4}
\begin{table}[t]
\caption{Wrapping numbers of the four O6-planes.} \vspace{0.4cm}
\begin{center}
\begin{tabular}{|c|c|c|}
\hline
  Orientifold Action & O6-Plane & $(n^1,l^1)\times (n^2,l^2)\times
(n^3,l^3)$\\
\hline
    $\Omega R$& 1 & $(2^{\beta_1},0)\times (2^{\beta_2},0)\times
(2^{\beta_3},0)$ \\
\hline
    $\Omega R\omega$& 2& $(2^{\beta_1},0)\times (0,-2^{\beta_2})\times
(0,2^{\beta_3})$ \\
\hline
    $\Omega R\theta\omega$& 3 & $(0,-2^{\beta_1})\times
(2^{\beta_2},0)\times
(0,2^{\beta_3})$ \\
\hline
    $\Omega R\theta$& 4 & $(0,-2^{\beta_1})\times (0,2^{\beta_2})\times
    (2^{\beta_3},0)$ \\
\hline
\end{tabular}
\end{center}
\label{orientifold}
\end{table}

In this paper, we concentrate on the supersymmetric AdS vacua with
metric, NSNS and RR fluxes~\cite{Camara:2005dc}.
For simplicity, we assume that the K\"ahler moduli $T_i$
satisfy $T_1=T_2=T_3$, then we obtain $q_1=q_2=q_3\equiv q$
from the superpotential in~\cite{Camara:2005dc}.
To satisfy the Jacobi identities
for metric fluxes, we consider the solution $a_i=a$,
$b_{ii}=-b_i$, and $b_{ji}=b_i$ in which $j\not= i$~\cite{Camara:2005dc}.

To have supersymmetric minima~\cite{Camara:2005dc}, we obtain that
\begin{eqnarray}
3 a {\rm Re}S~=~ b_i {\rm Re} U_i~, ~~{\rm for}~~ i=1, ~2, ~3~,~\,
\label{SUSY-min}
\end{eqnarray}
where
\begin{eqnarray}
{\rm Re}S~\equiv~{{e^{-\phi}}\over {\sqrt {\chi_1 \chi_2 \chi_3}}}~,~
{\rm Re}U_i~\equiv~e^{-\phi} {\sqrt {{\chi_j \chi_k}\over {\chi_i}}}~,~\,
\end{eqnarray}
where $S$ and $U_i$ are respectively
dilaton and complex structure moduli, $\phi$ is the four-dimensional
T-duality invariant dilaton, and $i \not= j \not= k \not= i$. And then we have
\begin{eqnarray}
b_1={{3a}\over {\chi_2 \chi_3}}~,~
b_2={{3a}\over {\chi_1 \chi_3}}~,~
b_3={{3a}\over {\chi_1 \chi_2}}~.~
\end{eqnarray}
Moreover, there are  consistency conditions
\begin{eqnarray}
3h_i a + h_0 b_i = 0~,~ ~{\rm for}~~ i=1,~2,~3~.~\,
\label{CONS-EQ}
\end{eqnarray}
So we have
\begin{eqnarray}
h_1=-{{h_0}\over {\chi_2 \chi_3}}~,~
h_2=-{{h_0}\over {\chi_1 \chi_3}}~,~
h_3=-{{h_0}\over {\chi_1 \chi_2}}~.~
\end{eqnarray}
Thus, the RR
tadpole cancellation conditions can be rewritten as following
\begin{eqnarray}
2^k N^{(1)} - \sum_a N_a A_a + {1\over 2}(h_0 m + 3 a q)& = & 16 ~,~\,
\label{N-tadodhz2z2I}
\end{eqnarray}
\begin{eqnarray}
-2^{\beta_1} N^{(2)} + \sum_a 2^{-\beta_2-\beta3} N_a B_a
- {1\over {2 \chi_2 \chi_3}} (h_0 m + 3 a q) & = & -2^{4-\beta_2-\beta_3} ~,~\,
\label{N-tadodhz2z2II}
\end{eqnarray}
\begin{eqnarray}
-2^{\beta_2} N^{(3)} +\sum_a 2^{-\beta_1-\beta_3} N_a C_a
- {1\over {2 \chi_1 \chi_3}} (h_0 m + 3 a q)  & = & - 2^{4-\beta_1-\beta_3} ~,~\,
\label{N-tadodhz2z2III}
\end{eqnarray}
\begin{eqnarray}
-2^{\beta_3} N^{(4)}
+\sum_a 2^{-\beta_1-\beta_2} N_a D_a - {1\over {2 \chi_1 \chi_2}} (h_0 m + 3 a q)
& = & - 2^{4-\beta_1-\beta_2} ~.~\,
\label{N-tadodhz2z2IV}
\end{eqnarray}
Therefore, if $(h_0 m + 3 a q) < 0$, the supergravity fluxes
contribute negative D6-brane charges to all the RR tadpole
cancellation conditions, and then, the RR tadpole cancellation
conditions give no constraints on the consistent model building
because we can always introduce suitable supergravity fluxes
and some stacks of D6-branes in
the hidden sector to cancel the RR tadpoles. Also,
if $(h_0 m + 3 a q) = 0$, the supergravity fluxes do not
contribute to any D6-brane charges, and then do not
affect the RR tadpole cancellation conditions.

In addition, the Freed-Witten anomaly cancellation
condition is~\cite{Camara:2005dc}
\begin{eqnarray}
-2^{-k} h_0 \tilde{A}_a + 2^{-\beta_1} h_1 \tilde{B}_a
+ 2^{-\beta_2} h_2 \tilde{C}_a
+ 2^{-\beta_3} h_3 \tilde{D}_a =0~.~\,
\end{eqnarray}
It can be shown that if Eqs. (\ref{susyconditions}),
(\ref{SUSY-min}), and (\ref{CONS-EQ}) are satisfied, the
 Freed-Witten anomaly is automatically cancelled.
So, we will not consider the Freed-Witten anomaly in 
our model building.

Furthermore, in addition to the above RR tadpole cancellation conditions,
the discrete D-brane RR
charges classified by $\mathbf{\Z_2}$ K-theory groups in the
presence of orientifolds, which are subtle and invisible by the
ordinary homology~\cite{MS,Witten9810188},
should also be taken into account~\cite{CU}. The K-theory conditions for a
$\mathbf{\Z_2\times \Z_2}$ orientifold are
\begin{equation}
\sum_a 2^{-k} \tilde{A}_a = \sum_a 2^{-\beta_1} \tilde{B}_a
  = \sum_a 2^{-\beta_2} \tilde{C}_a  = \sum_a 2^{-\beta_3} \tilde{D}_a  =
0 \textrm{ mod }4 \label{K-charges}~.~\,
\end{equation}

\subsection{Type IIA Theory on $\mathbf{T^6}$ Orientifold}

The intersecting D6-brane model building in Type IIA theory on
$\mathbf{T^6}$  orientifold
with flux compactifications is similar to that on
$\mathbf{T^6/(\Z_2\times \Z_2)}$  orientifold.
For the model building rules in the previous subsection,
we only need to make the following changes:
(1) For $a$ stack of $N_a$ D6-branes and its $\Omega R$ image, we have
$U(N_a)$ gauge symmetry, while for $a$ stack of $N_a$ D6-branes and its $\Omega R$ image
on the top of O6-plane, we obtain $USp(2 N_a)$ gauge symmetry. Also,
we present the  general spectrum of D6-branes' intersecting at generic angles
 in Type IIA theory on $\mathbf{T^6}$  orientifold
in Table \ref{Spectrum-T6}.
(2) We only have the $\Omega R$ O6-planes, so,
$[\Pi_{O6}]=[\Pi_{\Omega R}]$ in Eq. (\ref{intersections}), and the right-hand
sides of Eqs. (\ref{tadodhz2z2II}), (\ref{tadodhz2z2III})
and (\ref{tadodhz2z2IV}) are zero. (3) The metric, NSNS and RR
fluxes ($a_i$, $b_{ij}$, $h_0$, $h_i$, $m$ and $q_i$) are quantized in units of 2.
(4) To have three families of the SM fermions, we obtain that at least
one of the three two-tori is tilted. Thus,
the right-hand side of Eq. (\ref{K-charges})
is $0 \textrm{ mod }2$ for the K-theory conditions
in our model building~\cite{MarchesanoBuznego:2003hp}.

\begin{table}[h]
\renewcommand{\arraystretch}{1.5}
\center
\begin{tabular}{|c||c|}
\hline

Sector & Representation   \\ \hline \hline

$aa$ & $U(N_a )$ vector multiplet and 3 adjoint chiral multiplets \\
\hline

$ab+ba$ & $I_{ab}$ $ (N_a,
\overline{N_b})$ chiral multiplets  \\ \hline

$ab'+b'a$ & $I_{ab'}$ $ (N_a,
N_b)$ chiral multiplets  \\ \hline

$aa'+a'a$ & $  \frac 12 (I_{aa'} + I_{aO6}) $
anti-symmetric chiral multiplets  \\

  & $\frac 12 (I_{aa'} - I_{aO6})$
symmetric chiral multiplets  \\ \hline

\hline

\end{tabular}
\caption{The general spectrum for the intersecting D6-brane model
building in Type IIA theory on $\mathbf{T^6}$ orientifold
with flux compactifications, in particular,
$I_{aO6}=[\Pi_a][\Pi_{O6}]=-2^{3-k}l_a^1l_a^2l_a^3$.}
\label{Spectrum-T6}
\end{table}

\section{General Conditions for Pati-Salam Model Building}

In the $SU(5)$ models~\cite{Cvetic:2002pj}, 
flipped $SU(5)$ models~\cite{Chen:2005ab}, and trinification
models~\cite{Chen:2005mj}, the up-type quark Yukawa couplings, down-type quark Yukawa
couplings, and lepton Yukawa couplings are
forbidden by the anomalous $U(1)$ symmetries, respectively.
In the Pati-Salam models,  all
the SM fermions and Higgs fields belong to the bi-fundamental
representations, which can naturally arise from the intersecting
D6-brane model building. In particular, all the SM fermion
Yukawa couplings can in principle be allowed by the anomalous $U(1)$
symmetries. Therefore, the Pati-Salam models are very interesting
in the intersecting D6-brane model building.

In this paper, we construct the Pati-Salam models
with following properties:

\begin{itemize}

\item Three families of the SM fermions.

\item The Pati-Salam gauge symmetry can be broken down to
$SU(3)_C\times SU(2)_L\times U(1)_{B-L} \times U(1)_{I_{3R}}$ via D6-brane
splittings, and further down to the SM gauge symmetry around the string scale
via supersymmetry preserving Higgs mechanism.

\item The SM fermion Yukawa couplings are allowed by the anomalous $U(1)$
symmetries. We consider two kinds of Pati-Salam models: in the first
kind of models, we can give suitable masses to three families of the SM
fermions at the stringy tree level; in the second kind of models, we can
only give masses to the third family of the SM fermions
at tree level while we assume
that the masses for the first two families of the SM fermions
may be generated by quantum corrections.

\end{itemize}

To break the Pati-Salam gauge symmetry down to the
SM gauge symmetry via D6-brane splittings and supersymmetry preserving
Higgs mechanism, the $SU(4)_C$ and $SU(2)_R$ gauge symmetries
must come from $U(4)_C$ and $U(2)_R$ gauge symmetries
(see the following discussions).
Thus, we introduce three stacks of D6-branes, $a$, $b$, and $c$ with
the numbers of D6-branes 8, 4 (or 2), and 4 in
 Type IIA theory on $\mathbf{T^6/(\Z_2\times \Z_2)}$ orientifold, or
with the numbers of D6-branes 4, 2 (or 1), and 2 in
 Type IIA theory on $\mathbf{T^6}$ orientifold.
So, $a$, $b$, and $c$ stacks of D6-branes give
us the gauge symmetries $U(4)_C$, $U(2)_L$ (or $USp(2)_L$)
 and $U(2)_R$, respectively.
The anomalies from global $U(1)$s are cancelled by the generalized
 Green-Schwarz mechanism, and the gauge fields of these $U(1)$s
obtain masses via the linear $B\wedge F$ couplings. So, the
effective gauge symmetry is $SU(4)_C\times SU(2)_L\times SU(2)_R$.
In addition, we require that the intersection numbers satisfy
\begin{eqnarray}
\label{E3LF} I_{ab}~=~3~,~ I_{ab'}~=~0 ~{\rm if ~SU(2)_L ~from ~U(2)_L}~,~\,
\end{eqnarray}
\begin{eqnarray}
\label{E3RF} I_{ac} ~=~-3~,~ I_{ac'} ~=~0~.~\,
\end{eqnarray}
The conditions $I_{ab} =3$ and $I_{ac} =-3$ give us three
families of the SM fermions with quantum numbers $({\bf 4, 2, 1})$
and $({\bf {\bar 4}, 1, 2})$ under $SU(4)_C\times SU(2)_L\times
SU(2)_R$ gauge symmetry. $I_{ac'} =0 $ implies that $a$ stack of
D6-branes is parallel to the $\Omega R$ image $c'$
of the $c$ stack of D6-branes along at least one two-torus, for
example, the third two-torus. So, if $a$ and $c'$ stacks of
D6-branes are on the top of each other on the
third two-torus, we obtain $I_{ac'}^{(1,2)}$
pairs of the vector-like chiral multiplets with quantum
numbers $({\bf 4, 1, 2})$ and $({\bf {\bar 4}, 1, {\bar 2}})$,
where $I_{ac'}^{(1,2)}$ is the product of intersection numbers
for $a$ and $c'$ stacks of D6-branes on the first two two-tori.
 These particles are the Higgs fields which can break the
Pati-Salam gauge symmetry down to the SM gauge symmetry, and
preserve the D- and F-flatness, {\it i.~e.}, preserve supersymmetry.
Also, the conditions in Eq. (\ref{E3RF}) imply that the $SU(4)_C$ and $SU(2)_R$
gauge symmetries must come from $U(4)_C$ and $U(2)_R$ gauge symmetries,
respectively.

In addition, for the first kind of Pati-Salam models, we require that
\begin{eqnarray}
\label{E3H1} I_{bc} ~\ge~2~,~\,
\end{eqnarray}
and all the SM fermions and at least two bidoublet
Higgs fields arise from the intersections
on the same two-torus so that the suitable
three-family SM fermion masses and mixings can
be generated at the stringy tree level.
And for the second kind of Pati-Salam models, we require that
\begin{eqnarray}
\label{E3H2} I_{bc} ~\ge~1~,~\,
\end{eqnarray}
and the left-handed SM fermions, right-handed SM fermions
 and bidoublet Higgs fields do not arise from the intersections
on the same two-torus. Then we can only give
masses to the third family of the SM fermions at tree level, and we assume
that the first two families of the SM fermions can obtain masses
from quantum corrections.

In order to break the Pati-Salam gauge symmetry, we split the $a$ stack of
D6-branes into $a_1$ and $a_2$ stacks with respectively 6 (3) and 2 (1) D6-branes,
split the $c$ stack of D6-branes into $c_1$ and
$c_2$ stacks with 2 (1) D6-branes for each one
on Type IIA $\mathbf{T^6/(\Z_2\times \Z_2)}$ orientifold
(Type IIA $\mathbf{T^6}$ orientifold). And then, the
Pati-Salam gauge symmetry is broken down to
$SU(3)_C\times SU(2)_L\times U(1)_{B-L} \times U(1)_{I_{3R}}$. To
break this gauge symmetry down to the SM gauge symmetry, we assume
that the $a_2$ and $c_1'$ ($\Omega R$ image
of $c_1$) stacks of D6-branes are parallel and
on the top of each other on the third two-torus as
an example, and then we obtain $I_{a_2 c_1'}^{(1, 2)}$
 pairs of vector-like chiral multiplets
with quantum numbers $({\bf { 1}, 1, -1, 1/2})$ and $({\bf { 1},
1, 1, -1/2})$ under $SU(3)_C\times SU(2)_L\times U(1)_{B-L} \times
U(1)_{I_{3R}}$ gauge symmetry, where $I_{a_2 c_1'}^{(1, 2)}$ is
the product of intersection numbers for $a_2$ and $c_1'$ stacks on
the first two two-tori.
These particles can break the $SU(3)_C\times
SU(2)_L\times U(1)_{B-L} \times U(1)_{I_{3R}}$ gauge symmetry down to the SM
gauge symmetry and preserve supersymmetry in the mean time
 because their quantum numbers are the same as those of the right-handed neutrino and its
Hermitian conjugate. In summary, the complete
gauge symmetry breaking chains are
\begin{eqnarray}
SU(4)_C \times SU(2)_L \times SU(2)_R  &&
\overrightarrow{\;a\rightarrow a_1+a_2\;}\;  SU(3)_C\times SU(2)_L
\times SU(2)_R \times U(1)_{B-L} \nonumber\\&&
 \overrightarrow{\; c\rightarrow c_1+c_2 \;} \; SU(3)_C\times SU(2)_L\times
U(1)_{I_{3R}}\times U(1)_{B-L} \nonumber\\&&
 \overrightarrow{\;\rm Higgs \;
Mechanism\;} \; SU(3)_C\times SU(2)_L\times U(1)_Y~.~\,
\end{eqnarray}

\section{Two Kinds of Pati-Salam Models}

In this Section, we present the first and second kinds of Pati-Salam models
where all the SM fermion Yukawa couplings are allowed by the anomalous $U(1)$
symmetries.
Because the supergravity fluxes on Type IIA $\mathbf{T^6/(\Z_2\times \Z_2)}$
orientifold contribute large negative D6-brane charges due to
the Dirac quantization conditions
if we want to use them to relax the RR tadpole
cancellation conditions, many D6-branes in the hidden sector need to be
introduced so that the RR tadpoles can be completely cancelled.
 Then there may exist a lot of exotic particles. Therefore,
we mainly consider the Pati-Salam models on Type IIA
$\mathbf{T^6}$ orientifold with flux compactifications.
Also, we emphasize that the Pati-Salam models on Type IIA
 $\mathbf{T^6}$ orientifold
can be similarly constructed on Type IIA
 $\mathbf{T^6/(\Z_2\times \Z_2)}$ orientifold by introducing more
stacks of D6-branes in the hidden sector.
In addition, we determine the complex structure parameters via
 supersymmetric D6-brane configurations in  our model building.
Similar to~\cite{Camara:2005dc}, all the moduli may be stabilized
in our models.

\subsection{The First Kind of Pati-Salam Models}


\begin{table}[h]

\begin{center}
\small
\begin{tabular}{|c||@{}c@{}||@{}c@{}|@{}c@{}|@{}c@{}|} \hline

Representation & Multiplicity &$U(1)_a$&$U(1)_b$&$U(1)_c$   \\
\hline \hline

$(4_a,\bar{2}_b)$ & 3 & 1 & -1 & 0    \\

$(\bar{4}_a,2_c)$ & 3 & -1 & 0 & 1     \\

$(2_b,\bar{2}_c)$ & 6 & 0 & 1 & -1     \\

\hline

$(4_a,2_c)$ & 3 & 1 & 0 & 1  \\

$(\bar{4}_a,\bar{2}_c)$ & 3 & -1 & 0 & -1   \\ \hline \hline

$6_a$ & 1 & 2 & 0 & 0  \\

$\overline{10}_a$ & 1 & -2 & 0 & 0  \\

$1_c$ & 2 & 0 & 0 & -2  \\

$3_c$ & 2 & 0 & 0 & 2 \\ \hline

\end{tabular}
\caption{The particle spectrum in observable sector in Model TI-U-4 with
gauge symmetry $[U(4)_C\times U(2)_L\times
U(2)_R]_{observable} \times [U(2)\times USp(2)\times
USp(10)]_{hidden}$. Here,
$a$, $b$ and $c$ denote the
gauge groups $U(4)_C$, $U(2)_L$ and $U(2)_R$, respectively. }
\label{TI-U-4-spectrum-A}
\end{center}
\end{table}

\begin{table}[h]

\begin{center}
\small
\begin{tabular}{|c||@{}c@{}||@{}c@{}|@{}c@{}|@{}c@{}|@{}c@{}|} \hline

Representation & Multiplicity &$U(1)_a$&$U(1)_b$&$U(1)_c$& $U(1)_d$   \\
\hline \hline

$(4_a,\bar{2}_d)$ & 2 & 1 & 0 & 0 & -1    \\

$({\bar 4}_a,2_e)$ & 3 & -1 & 0 & 0 & 0 \\

$(4_a, 10_{O6})$ & 1 & 1 & 0 & 0 & 0    \\

$(\bar{2}_b,2_d)$ & 1 & 0 & -1 & 0 & 1    \\

$(\bar{2}_b,\bar{2}_d)$ & 5 & 0 & -1 & 0 & -1   \\

$(2_b,2_e)$ & 6 & 0 & 1 & 0 & 0    \\

$(2_c,2_d)$       & 2 & 0 & 0 & 1 & 1   \\

$({\bar 2}_c,10_{O6})$       & 2 & 0 & 0 & -1 & 0   \\

$(2_d, 2_e)$ & 6 & 0 & 0 & 0 & 1 \\ \hline

\end{tabular}
\caption{The exotic particle spectrum in Model TI-U-4 with
gauge symmetry $[U(4)_C\times U(2)_L\times
U(2)_R]_{observable} \times [U(2)\times USp(2)\times
USp(10)]_{hidden}$. Here,
$a$, $b$, $c$, $d$, $e$ and $O6$ denote the
gauge groups $U(4)_C$, $U(2)_L$, $U(2)_R$, $U(2)$,
$USp(2)$ and $USp(10)$, respectively. }
\label{TI-U-4-spectrum-B}
\end{center}
\end{table}


We present the D6-brane configurations and intersection numbers for
the first kind of Pati-Salam models, {\it i.~e.}, the Models TI-U-i with
i=1, ..., 7, and the Models TI-Sp-j with j=1, ..., 4,
 in Tables \ref{TI-U-1}-\ref{TI-Sp-4}.
In these models, all the SM fermions and at least three bidoublet
Higgs fields arise from the intersections on the same two-torus.
So, the suitable three-family
SM fermion masses and mixings can be generated
at the stringy tree level, and then the rank one
problem for the SM fermion Yukawa matrices can
be solved.

The observable gauge symmetry in Models TI-U-i
is $U(4)_C\times U(2)_L\times U(2)_R$. 
The Models TI-U-i with i=1, ..., 5 are  on Type IIA
$\mathbf{T^6}$ orientifold while the Models TI-U-6
and TI-U-7 are on Type IIA $\mathbf{T^6/(\Z_2\times \Z_2)}$
orientifold. Only Model TI-U-4 has a $U(4)_C$ symmetric
representation. Moreover, there are six bidoublet Higgs fields
in Models  TI-U-1, TI-U-2, TI-U-3 and TI-U-4.
There are twelve pairs of vector-like bidoublet Higgs fields
from the massless open string states in a $N=2$ subsector
in  Model TI-U-5, and six pairs of vector-like bidoublet Higgs fields
in Models TI-U-6 and TI-U-7. Especially, the
D6-brane configurations in  Model TI-U-6 are the same
as those in Model I-Z-10 in Ref.~\cite{CLL},
and Model TI-U-6 is the only model that the
supergravity fluxes do not contribute to the
D6-brane RR tadpoles. Also, the D6-brane configurations in
the observable sector in  Model TI-U-7 are the same
as those in Model I-Z-10 in Ref.~\cite{CLL},
and there are a lot of exotic particles
from extra gauge  groups due to the large supergravity fluxes
 and the RR tadpole cancellation conditions.

The observable gauge symmetry in Models TI-Sp-j
is $U(4)_C\times USp(2)_L\times U(2)_R$, and
all these models are  on Type IIA
$\mathbf{T^6}$ orientifold. There are $U(4)_C$ symmetric
representations in Models TI-Sp-3 and
TI-Sp-4. Also, there are three bidoublet Higgs fields
in Models TI-Sp-1, TI-Sp-2 and TI-Sp-3, and three pairs of
vector-like bidoublet Higgs fields
in Model TI-Sp-4.

We present the complete  particle spectrum in Model TI-U-4
with six bidoublet Higgs fields
in Tables \ref{TI-U-4-spectrum-A} and \ref{TI-U-4-spectrum-B},
and the particle spectrum in the observable sector in Model TI-Sp-1
with three bidoublet Higgs fields
in Table \ref{TI-Sp-1-spectrum}.
The vector-like particles with quantum numbers
 $(4_a,2_c)$ and $(\bar{4}_a,\bar{2}_c)$ are the
Higgs fields which can break the Pati-Salam gauge symmetry
down to the SM gauge symmetry. After suitable D6-brane splittings,
only the vector-like particles with quantum numbers
 $({\bf { 1}, 1, -1, 1/2})$ and $({\bf { 1},
1, 1, -1/2})$ under $SU(3)_C\times SU(2)_L\times U(1)_{B-L} \times
U(1)_{I_{3R}}$ gauge symmetry from $(4_a,2_c)$ and $(\bar{4}_a,\bar{2}_c)$
are massless, and they can break the
$U(1)_{B-L} \times U(1)_{I_{3R}}$ gauge symmetry down
to the $U(1)_Y$ gauge symmetry by supersymmetry preserving
Higgs mechanism. It is apparent from Tables \ref{TI-U-4-spectrum-A}
and \ref{TI-Sp-1-spectrum} that all the SM fermion Yukawa couplings
are allowed by the anomalous $U(1)_{a, b, c}$ symmetries.


\begin{table}[h]

\begin{center}
\small
\begin{tabular}{|c||@{}c@{}||@{}c@{}|@{}c@{}|@{}c@{}|} \hline

Representation & Multiplicity & $U(1)_a$ & $U(1)_b$ & $U(1)_c$ \\ \hline \hline

$(4_a,\bar{2}_b)$ & 3 & 1 & -1 & 0 \\

$(\bar{4}_a,2_c)$ & 3 & -1 & 0 & 1 \\

$(2_b,\bar{2}_c)$ & 3 & 0 & 1 & -1 \\

\hline

$(4_a,2_c)$ & 1 & 1 & 0 & 1 \\

$(\bar{4}_a,\bar{2}_c)$ & 1 & -1 & 0 & -1
\\ \hline \hline
\multicolumn{5}{|c|}{Exotic Particles and Hidden Sector Matter}\\
\hline
\end{tabular}
\caption{The particle spectrum in the observable sector in Model TI-Sp-1
with gauge symmetry $[U(4)_C\times USp(2)_L\times
U(2)_R]_{observable} \times [U(2)\times U(1)^4 \times
USp(2)]_{hidden}$. Here,
$a$, $b$ and $c$ denote
the gauge groups $U(4)_C$, $USp(2)_L$ and $U(2)_R$, respectively. }
\label{TI-Sp-1-spectrum}
\end{center}
\end{table}


\subsection{The Second Kind of Pati-Salam Models}

We present the D6-brane configurations and intersection numbers for
the second kind of Pati-Salam models, {\it i.~e.}, the Models TII-U-i with
i=1, ..., 6, and the Models TII-Sp-j with j=1, ..., 5,
 in Tables \ref{TII-U-1}-\ref{TII-Sp-4}.
Because the left-handed SM fermions,
 right-handed SM fermions and bidoublet Higgs fields
 do not arise from the intersections
on the same two-torus in these models, only the SM fermion masses
for the third family can be generated
at the stringy tree level. We assume that the first two families of the SM fermions
may obtain masses from quantum corrections.

Similar to the above subsection, the observable gauge symmetry in Models
TII-U-i is $U(4)_C\times U(2)_L\times U(2)_R$. 
All these models are  on Type IIA
$\mathbf{T^6}$ orientifold. Moreover,
there are two bidoublet Higgs fields
in Model TII-U-1, three in Model  TII-U-2,
and four in Models TII-U-3 and TII-U-4.
There are two and four pairs of
vector-like bidoublet Higgs fields
in Models TII-U-5 and TII-U-6, respectively.

The observable gauge symmetry in Models TII-Sp-j
is $U(4)_C\times USp(2)_L\times U(2)_R$. 
The Models TII-Sp-i with i=1, ..., 4 are  on Type IIA
$\mathbf{T^6}$ orientifold while the Model TII-Sp-5
is on Type IIA $\mathbf{T^6/(\Z_2\times \Z_2)}$
orientifold. Also, there are $U(4)_C$ symmetric
representations in Models TII-Sp-1, TII-Sp-3 and
TII-Sp-4. Moreover, there are one bidoublet Higgs field
in Model TII-Sp-1, three in Models
 TII-Sp-2, TII-Sp-3 and TII-Sp-5, and four pairs of
vector-like bidoublet Higgs fields
in Model TII-Sp-4.

\section{Discussion and Conclusions}

 We considered the Pati-Salam model building on Type IIA orientifolds
with flux compactifications in supersymmetric AdS vacua.
We showed that the metric, NSNS and RR fluxes can contribute
negative D6-brane charges to all the RR tadpole cancellation conditions,
{\it i.~e.}, the RR tadpole cancellation conditions can
be completely relaxed. So, the major constraints on
  consistent model building on Type IIA orientifolds
are the four-dimensional $N=1$ supersymmetry conditions.

We constructed two kinds of three-family Pati-Salam models.
In the first kind of models, the suitable three-family
SM fermion masses and mixings can be generated at the 
stringy tree level, and then the rank one problem for the SM
fermion Yukawa matrices can be solved. While in the second kind
of models, only the third-family SM fermions can have
masses at tree level, and we assume that the first two
families of the SM fermions may obtain masses from quantum corrections.
In these models, the complex structure parameters
can be determined by supersymmetric
D6-brane configurations, and all the moduli may be stabilized.
The initial gauge symmetries $U(4)_C \times U(2)_L \times U(2)_R$ and
$U(4)_C \times USp(2)_L \times U(2)_R$ can be
broken down to the $SU(3)_C\times SU(2)_L\times U(1)_{B-L} \times
U(1)_{I_{3R}}$ due to  the generalized Green-Schwarz mechanism
and  D6-brane splittings, and further down to the
SM gauge symmetry at about the string scale via the supersymmetry
preserving Higgs mechanism.
 Moreover, most of our models are on Type IIA
$\mathbf{T^6}$ orientifold with flux compactifications
because the supergravity fluxes on Type IIA $\mathbf{T^6/(\Z_2\times \Z_2)}$
orientifold contribute large negative D6-brane RR tadpoles
 due to the Dirac quantization conditions if we use them to
relax the RR tadpole cancellation conditions.
Of course, our Pati-Salam models on Type IIA $\mathbf{T^6}$ orientifold
can be similarly constructed on Type IIA
 $\mathbf{T^6/(\Z_2\times \Z_2)}$ orientifold by introducing more
stacks of D6-branes in the hidden sector.

For the first kind of Pati-Salam
models with $U(4)_C\times U(2)_L\times U(2)_R$
gauge symmetry, we presented five models on Type IIA
$\mathbf{T^6}$ orientifold and two models on
Type IIA $\mathbf{T^6/(\Z_2\times \Z_2)}$ orientifold.
We have  six bidoublet Higgs fields
in the Models TI-U-i with
i=1, ..., 4, twelve pairs of vector-like bidoublet Higgs fields
in  Model TI-U-5, and six pairs of vector-like bidoublet Higgs fields
in Models TI-U-6 and TI-U-7. Especially, Model TI-U-6 is the only
model that its supergravity  fluxes do not contribute to the
D6-brane RR tadpoles.
Moreover, for the first kind of Pati-Salam models with
$U(4)_C\times USp(2)_L\times U(2)_R$ gauge symmetry, we
gave four models on Type IIA
$\mathbf{T^6}$ orientifold. Models TI-Sp-1, TI-Sp-2 and TI-Sp-3
have three bidoublet Higgs fields, and Model TI-Sp-4
has three pairs of vector-like bidoublet Higgs fields.

For the second kind of Pati-Salam models with
 $U(4)_C\times U(2)_L\times U(2)_R$
gauge symmetry, we constructed six models on Type IIA
$\mathbf{T^6}$ orientifold. There are  two bidoublet Higgs fields
in Model TII-U-1, three in
Model  TII-U-2, and four in Models TII-U-3 and TII-U-4.
We also have two and four pairs of
vector-like bidoublet Higgs fields
in Models TII-U-5 and  TII-U-6, respectively.
For the second kind of Pati-Salam models with
$U(4)_C\times USp(2)_L\times U(2)_R$ gauge symmetry,
we presented four models on Type IIA
$\mathbf{T^6}$ orientifold, and one model
on Type IIA $\mathbf{T^6/(\Z_2\times \Z_2)}$
orientifold. There are one bidoublet Higgs field
in  Model TII-Sp-1, three in Models
 TII-Sp-2, TII-Sp-3 and TII-Sp-5, and four pairs of
vector-like bidoublet Higgs fields in Model TII-Sp-4.
Comparing to the previous Pati-Salam model building
on Type IIA $\mathbf{T^6/(\Z_2\times \Z_2)}$
orientifold without supergravity fluxes, we
have less bidoublet Higgs fields.

Furthermore, the generic feature of our models is that
there exist some exotic particles, in particular, in
Models TI-U-7 and TII-Sp-5 because their supergravity fluxes
contribute large negative D6-brane charges
and we have to introduce quite a few stacks of D6-branes
in the hidden sector to cancel the RR tadpoles. The phenomenological
consequences of our models, for instance, the SM fermion masses and
mixings, the moduli stabilization, and how to give
masses to exotic particles, will be presented
in detail elsewhere~\cite{CLN}.

\section*{Acknowledgments}
The research of T.L. was supported by DOE grant DE-FG02-96ER40959,
and the research of D.V.N. was supported by DOE grant
DE-FG03-95-Er-40917.

\newpage

\appendix

\section{ The D6-Brane Configurations and Intersection Numbers
for  the First Kind of Pati-Salam Models}
In  Appendix A, we present the D6-brane configurations and
intersection numbers
for  the first kind of Pati-Salam models. Let us explain the convention.
Suppose $b$ and $c$ stacks of D6-branes are parallel on
a two-torus and the product of intersection numbers on
the other two two-tori is $i$, we denote their
intersection number as $0(i)$.

\subsection{$U(4)_C\times U(2)_L \times U(2)_R$ Models}

\begin{table}[h]
\begin{center}
\footnotesize
\begin{tabular}{|@{}c@{}|c||@{}c@{}c@{}c@{}||c|c||c|@{}c@{}|c|@{}c@{}||
@{}c@{}|@{}c@{}|@{}c@{}|@{}c@{}|@{}c@{}|c|@{}c@{}|} \hline stack &
$N$ & ($n_1$,$l_1$) & ($n_2$,$l_2$) & ($n_3$,$l_3$) & A & S & $b$
& $b'$ & $c$ & $c'$ & $d$ & $d'$ & $e$ & $e'$ & $f$ & $f'$ & $O6$
  \\ \hline \hline

$a$ & 4 & ( 1, 0) & (-1,-1) & (-1, 1) & 0 & 0 & 3 & 0(3) & -3 &

0(3) & 0(3) &-3 & 0(3) & 3 & 0(1) & - & 0(1)

\\ \hline

$b$ & 2 & ( 1,-3) & ( 1,-1) & ( 0, 2) & -6 & 6 & - & - & 6 & 0(1)

& -6 & 0(18) & -9 & -3 & 0(3) & - & -6

\\ \hline

$c$ & 2 & (-1,-3) & ( 0, 2) & ( 1, 1) & 6 & -6 & - & - & - & - & 9

& 3 & 6 & 0(18) & 0(3) & - & 6

\\ \hline \hline

$d$ & 2 & ( 2,-3) & ( 1, 1) & ( 2, 0) & 0 & 0 & - & - & - & - & -

& - & 12 & 0(1) & -6 & - & 0(3)

\\ \hline

$e$ & 2 & ( 2, 3) & ( 2, 0) & ( 1,-1) & 0 & 0 & - & - & - & - & -

& - & - & - & 6 & - & 0(3)

\\ \hline

$f$ & 1 & ( 1, 0) & ( 0,-2) & ( 0, 2) & 0 & 0 & - & - & - & - & -

& - & - & - & - & - & 0(4)

\\ \hline

$O6$ & 1 & ( 1, 0) & ( 2, 0) & ( 2, 0) & - & - & - & - & - & - & -

& - & - & - & - & - & -

\\ \hline

\end{tabular}

\caption{D6-brane configurations and intersection numbers
for Model TI-U-1 on Type IIA $\mathbf{T^6}$ orientifold.
The complete gauge symmetry is
$[U(4)_C \times U(2)_L \times U(2)_R]_{observable}\times
[U(2)^2 \times USp(2)^2]_{hidden}$, the
SM fermions and Higgs fields arise from
the intersections on the first two-torus,
and the complex structure parameters are
$3 \chi_1 = \chi_2 = \chi_3=\sqrt{2}$.
To satisfy the RR tadpole cancellation conditions,
we choose $h_0=-2(3q+4)$, $a=4$, and $m=2$.}
\label{TI-U-1}
\end{center}
\end{table}

\newpage


\begin{table}[h]

\begin{center}

\footnotesize

\begin{tabular}{|@{}c@{}|c||@{}c@{}c@{}c@{}||c|c||c|@{}c@{}|c|@{}c@{}||
@{}c@{}|@{}c@{}|@{}c@{}|@{}c@{}|@{}c@{}|@{}c@{}|@{}c@{}|@{}c@{}|@{}c@{}|c|@{}c@{}|c|}

\hline

stack & $N$ & ($n_1$,$l_1$) & ($n_2$,$l_2$) & ($n_3$,$l_3$) & A & S

& $b$ & $b'$ & $c$ & $c'$ & $d$ & $d'$ & $e$ & $e'$ & $f$ & $f'$ &

$g$ &$g'$ & $h$ & $h'$ & $i$ & $i'$   \\ \hline \hline

$a$ & 4 & ( 1, 0) & (-1,-1) & (-1, 1) & 0 & 0 & 3 & 0(3) & -3 &

0(3) & -1 & 2 & 1 & -2 & 3 & 0(3) & -3 & 0(3) & 1 & - & -1 & -

\\ \hline

$b$ & 2 & (-1, 3) & (-2, 0) & ( 1, 1) & 0 & 0 & - & - & 6 & 0(1) &

0(8) & 0(2) & -2 & 4 & 0(1) & -6 & 0(12) & 0(2) & 0(1) & - & 2 & -

\\ \hline

$c$ & 2 & ( 1, 3) & (-1, 1) & (-2, 0) & 0 & 0 & - & - & - & - & 2

& -4 & 0(8) & 0(2) & 0(12) & 0(2) & 0(1) & 6 & -2 & - & 0(1) & -   \\

\hline \hline

$d$ & 1 & ( 1, 1) & ( 2, 0) & ( 3,-1) & 0 & 0 & - & - & - & - & -

& - & -2 & 0(1) & 12 & -6 & -8 & 8 & 0(3) & - & -2 & -  \\

\hline

$e$ & 1 & ( 1,-1) & ( 3, 1) & ( 2, 0) & 0 & 0 & - & - & - & - & -

& - & - & - & 8 & -8 & -12 & 6 & 2 & - & 0(3) & -   \\ \hline

$f$ & 1 & ( 1,-3) & ( 1,-1) & ( 0, 2) & -6 & 6 & - & - & - & - & -

& - & - & - & - & - & 6 & 0(1) & 0(1) & - & 2 & -   \\ \hline

$g$ & 1 & ( 1, 3) & ( 0,-2) & ( 1, 1) & 6 & -6 & - & - & - & - & -

& - & - & - & - & - & - & - & -2 & - & 0(1) & -  \\ \hline

$h$ & 1 & ( 0,-1) & ( 2, 0) & ( 0, 2) & 0 & 0 & - & - & - & - & -

& - & - & - & - & - & - & - & - & - & 0(4) & -  \\ \hline

$i$ & 1 & ( 0,-1) & ( 0, 2) & ( 2, 0) & 0 & 0 & - & - & - & - & -

& - & - & - & - & - & - & - & - & - & - & -  \\ \hline

\end{tabular}

\caption{D6-brane configurations and intersection numbers
for Model TI-U-2 on Type IIA $\mathbf{T^6}$ orientifold.
The complete gauge symmetry is
$[U(4)_C \times U(2)_L \times U(2)_R]_{observable}\times
[U(1)^4 \times USp(2)^2]_{hidden}$, the
SM fermions and Higgs fields arise from
the intersections on the first two-torus,
and the complex structure parameters are
$6\chi_1=\chi_2=\chi_3=2$.
To satisfy the RR tadpole cancellation conditions,
we choose $h_0=-4(3q+2)$, $m=2$, and $a=8$.}
\label{TI-U-2}
\end{center}
\end{table}

\newpage


\begin{table}[h]

\begin{center}

\footnotesize

\begin{tabular}{|@{}c@{}|c||@{}c@{}c@{}c@{}||c|c||c|@{}c@{}|c|@{}c@{}||
@{}c@{}|@{}c@{}|@{}c@{}|@{}c@{}|@{}c@{}|@{}c@{}|@{}c@{}|@{}c@{}|@{}c@{}|c|@{}c@{}|c|}

\hline

stack & $N$ & ($n_1$,$l_1$) & ($n_2$,$l_2$) & ($n_3$,$l_3$) & A & S

& $b$ & $b'$ & $c$ & $c'$ & $d$ & $d'$ & $e$ & $e'$ & $f$ & $f'$

& $g$ & $g'$ & $h$ & $h'$ & $i$ & $i'$  \\

\hline \hline

$a$ & 4 & ( 1, 1) & ( 1,-3) & ( 1, 0) & 0 & 0 & 3 & 0(1) & -3 &

0(3) & 0(3) & 0(1) & 0(4) & 0(1) & 0(1) & -3 & -1 & 2 & 1 & - &

0(1) & - \\ \hline

$b$ & 2 & ( 2, 0) & ( 1, 3) & ( 1,-1) & 0 & 0 & - & - & 6 & 0(3) &

0(1) & 3 & 1 & -2 & 0(6) & 0(4) & 0(8) & 0(2) & 0(1) & - & -2 & -  \\

\hline

$c$ & 2 & ( 1,-1) & ( 2, 0) & ( 1, 1) & 0 & 0 & - & - & - & - & -3

& 0(3) & -1 & 2 & 6 & 0(3) & 2 & -4 & -2 & - & 2 & -

\\ \hline  \hline

$d$ & 2 & ( 1, 1) & (-1,-3) & ( 0, 1) & 3 & -3 & - & - & - & - & -

& - & 2 & 2 & -3 & 0(1) & 6 & 3 & 0(1) & - & -1 & -

\\ \hline

$e$ & 2 & ( 3,-1) & ( 1, 1) & ( 1, 0) & 0 & 0 & - & - & - & - & -

& - & - & - & 6 & -3 & -1 & 0(1) & -1 & - & 0(3) & -

\\ \hline

$f$ & 1 & ( 0, 2) & (-1, 3) & (-1, 1) & -6 & 6 & - & - & - & - & -

& - & - & - & - & - & -8 & -8 & 2 & - & 0(1) & -

\\ \hline

$g$ & 1 & ( 2, 0) & ( 1,-1) & ( 3, 1) & 0 & 0 & - & - & - & - & -

& - & - & - & - & - & - & - & 0(3) & - & 2 & -

\\ \hline

$h$ & 1 & ( 2, 0) & ( 0,-2) & ( 0, 1) & 0 & 0 & - & - & - & - & -

& - & - & - & - & - & - & - & - & - & 0(2) & -

\\ \hline

$i$ & 1 & ( 0,-2) & ( 0, 2) & ( 1, 0) & 0 & 0 & - & - & - & - & -

& - & - & - & - & - & - & - & - & - & - & -

\\ \hline

\end{tabular}
\caption{D6-brane configurations and intersection numbers
for Model TI-U-3 on Type IIA $\mathbf{T^6}$ orientifold.
The complete gauge symmetry is
$[U(4)_C \times U(2)_L \times U(2)_R]_{observable} \times
[U(2)^2 \times U(1)^2 \times USp(2)^2]_{hidden}$, the
SM fermions and Higgs fields arise from
the intersections on the second two-torus,
and the complex structure parameters are
$\chi_1 = 3\chi_2 = 2\chi_3 = 2$.
To satisfy the RR tadpole cancellation conditions,
we choose $h_0 = -4(3q+2)$, $m=2$, and $a=8$.}
\label{TI-U-3}
\end{center}
\end{table}

\newpage



\begin{table}[h]

\begin{center}

\footnotesize

\begin{tabular}{|@{}c@{}|c||@{}c@{}c@{}c@{}||c|c||c|@{}c@{}|c|@{}c@{}||
@{}c@{}|@{}c@{}|@{}c@{}|c|@{}c@{}|} \hline

stack & $N$ & ($n_1$,$l_1$) & ($n_2$,$l_2$) & ($n_3$,$l_3$) & A &

S & $b$ & $b'$ & $c$ & $c'$ & $d$ & $d'$ & $e$ & $e'$ & $O6$

  \\ \hline \hline

$a$ & 4 & ( 0,-1) & ( 1, 1) & ( 3, 1) & 1 & -1 & 3 & 0(1) & -3 &

0(3) & 2 & 0(2) & -3 & - & 1

\\ \hline

$b$ & 2 & (-1,-1) & ( 2, 0) & (-3, 1) & 0 & 0 & - & - & 6 & 0(3) &

-1 & -5 & 6 & - & 0(1)

\\ \hline

$c$ & 2 & ( 1,-1) & (-1, 1) & ( 0,-2) & -2 & 2 & - & - & - & - &

0(10) & 2 & 0(1) & - & -2

\\ \hline \hline

$d$ & 2 & ( 2, 3) & ( 1,-1) & ( 2, 0) & 0 & 0 & - & - & - & - & -

& - & 6 & - & 0(3)

\\ \hline

$e$ & 1 & ( 1, 0) & ( 0,-2) & ( 0, 2) & 0 & 0 & - & - & - & - & -

& - & - & - & 0(4)

\\ \hline

$O6$ & 5 & ( 1, 0) & ( 2, 0) & ( 2, 0) & - & - & - & - & - & - & -

& - & - & - & -

\\ \hline

\end{tabular}
\caption{D6-brane configurations and intersection numbers
for Model TI-U-4 on Type IIA $\mathbf{T^6}$ orientifold.
The complete gauge symmetry is
$[U(4)_C \times U(2)_L \times U(2)_R]_{observable}\times
[U(2) \times USp(2) \times USp(10)]_{hidden}$, the
SM fermions and Higgs fields arise from
the intersections on the third two-torus,
and the complex structure parameters are
$6\chi_1= 2\chi_2= \chi_3= 2\sqrt{6}$.
To satisfy the RR tadpole cancellation conditions,
we choose $h_0=-12(q+2)$, $m=2$, and $a=8$.}
\label{TI-U-4}
\end{center}
\end{table}

\newpage


\begin{table}[h]

\begin{center}

\footnotesize

\begin{tabular}{|@{}c@{}|c||@{}c@{}c@{}c@{}||c|c||c|@{}c@{}|@{}c@{}|@{}c@{}||
@{}c@{}|@{}c@{}|@{}c@{}|@{}c@{}|@{}c@{}|@{}c@{}|@{}c@{}|} \hline

stack & $N$ & ($n_1$,$l_1$) & ($n_2$,$l_2$) & ($n_3$,$l_3$) & A &

S & $b$ & $b'$ & $c$ & $c'$ &

$d$ & $d'$ & $e$ & $e'$ & $f$ & $f'$ & $O6$    \\

\hline \hline

$a$ & 4 & ( 1, 0) & (-1,-1) & (-1, 1) & 0 & 0 & 3 & 0(3) & -3 &

0(3) & 0(1) & 0(1) & 2 & 2 & -2 & -2 & 0(1)

\\ \hline

$b$ & 2 & ( 1,-3) & ( 1,-1) & ( 0, 2) & -6 & 6 & - & - & 0(12) &

0(2) & 0(1) & 1 & 2 & -1 & 6 & 3 & -6  \\ \hline

$c$ & 2 & ( 1, 3) & (-1, 1) & (-2, 0) & 0 & 0 & - & - & - & - &

0(1) & -1 & -6 & -3 & -2 & 1 & 0(3)

\\ \hline  \hline

$d$ & 2 & ( 0, 1) & ( 1,-1) & ( 1,-1) & -1 & 1 & - & - & - & - & -

& - & 0(4) & 0(1) & 0(4) & 0(1) & -1

\\ \hline

$e$ & 1 & ( 0,-1) & ( 3, 1) & ( 1, 3) & 3 & -3 & - & - & - & - & -

& - & - & - & 0(16) & 0(25) & 3

\\ \hline

$f$ & 1 & ( 0,-1) & ( 1, 3) & ( 3, 1) & 3 & -3 & - & - & - & - & -

& - & - & - & - & - & 3

\\ \hline

$O6$ & 4 & ( 1, 0) & ( 2, 0) & ( 2, 0) & - & - & - & - & - & - & -

& - & - & - & - & - & - \\ \hline

\end{tabular}

\caption{D6-brane configurations and intersection numbers
for Model TI-U-5 on Type IIA $\mathbf{T^6}$ orientifold.
The complete gauge symmetry is
$[U(4)_C \times U(2)_L \times U(2)_R]_{observable} \times
[U(2) \times U(1)^2 \times USp(8)]_{hidden}$, the
SM fermions and Higgs fields arise from
the intersections on the first two-torus,
and the complex structure parameters are
$6\chi_1 = \chi_2 = \chi_3 = 2$.
To satisfy the RR tadpole cancellation conditions,
we choose $h_0 = -4(3q+2)$, $m=2$, and $a=8$.}
\label{TI-U-5}
\end{center}
\end{table}

\newpage


\begin{table}[h]

\begin{center}

\footnotesize

\begin{tabular}{|@{}c@{}|c||@{}c@{}c@{}c@{}||c|c||c|@{}c@{}|@{}c@{}|
@{}c@{}||@{}c@{}|@{}c@{}|@{}c@{}|@{}c@{}|} \hline

stack & $N$ & ($n_1$,$l_1$) & ($n_2$,$l_2$) & ($n_3$,$l_3$) & A & S

& $b$ & $b'$ & $c$ & $c'$ & $O6^{1}$ & $O6^{2}$ & $O6^{3}$ & $O6^{4}$ \\ \hline \hline

$a$ & 8 & ( 0,-1) & ( 1, 1) & ( 1, 1) & 0 & 0 & 3 & 0(3) & -3 &

0(3) & 1 & -1 & 0 & 0 \\ \hline

$b$ & 4 & ( 3, 1) & ( 1, 0) & ( 1,-1) & -2 & 2 & - & - & 0(6) &

0(1) & 0 & 1 & 0 & -3  \\  \hline

$c$ & 4 & ( 3,-1) & ( 0, 1) & ( 1,-1) & 2 & -2 & - & - & - & - &
-1 & 0 & 3 & 0 \\ \hline \hline

$O6^{1}$ & 2 & ( 1, 0) & ( 1, 0) & ( 2, 0) & - & - & - & - & - & -

& - & - & - & -  \\ \hline

$O6^{2}$ & 2 & ( 1, 0) & ( 0,-1) & ( 0, 2) & - & - & - & - & - & -

& - & - & - & -  \\ \hline

$O6^{3}$ & 2 & ( 0, -1) & ( 1, 0) & ( 0, 2) & - & - & - & - & - & -

& - & - & - & -  \\ \hline

$O6^{4}$ & 2 & ( 0, -1) & ( 0, 1) & ( 2, 0) & - & - & - & - & - & -

& - & - & - & -  \\ \hline
\end{tabular}
\caption{D6-brane configurations and intersection numbers
for Model TI-U-6 on Type IIA $\mathbf{T}^6 / \Z_2 \times \Z_2$ orientifold.
The complete gauge symmetry is
$[U(4)_C \times U(2)_L \times U(2)_R]_{observable}\times
[ USp(2)^4]_{hidden}$, the
SM fermions and Higgs fields arise from
the intersections on the first two-torus,
and the complex structure parameters are
$2\chi_1=6\chi_2=3\chi_3 =6$.
To satisfy the RR tadpole cancellation conditions,
we choose $h_0=\pm 3a$,  and $m=\mp q$
so that the supergravity
fluxes do not contribute to the D6-brane RR tadpoles.}
\label{TI-U-6}
\end{center}
\end{table}

\newpage


\begin{table}[h]

\begin{center}

\footnotesize

\begin{tabular}{|@{}c@{}|c||@{}c@{}c@{}c@{}||c|c||c|@{}c@{}|@{}c@{}|
@{}c@{}||@{}c@{}|@{}c@{}|@{}c@{}|@{}c@{}|@{}c@{}|@{}c@{}|@{}c@{}|
@{}c@{}|@{}c@{}|@{}c@{}|@{}c@{}|@{}c@{}|@{}c@{}|@{}c@{}|} \hline

stack & $N$ & ($n_1$,$l_1$) & ($n_2$,$l_2$) & ($n_3$,$l_3$) & A & S

& $b$ & $b'$ & $c$ & $c'$ & $d$ & $d'$ & $e$ & $e'$ & $f$ & $f'$ &

$g$ & $g'$ & $h$ & $h'$ & $i$ & $i'$ & $O6^{1}$ & $O6^{2}$

\\ \hline \hline

$a$ & 8 & ( 0,-1) & ( 1, 1) & ( 1, 1) & 0 & 0 & 3 & 0(3) & -3 &

0(3) & 18 & 36 & 15 & -9 & -18 & -36 & -15 & 9 & 8 & 8 & -8 & -8 &

1 & -1   \\ \hline

$b$ & 4 & ( 3, 1) & ( 1, 0) & ( 1,-1) & -2 & 2 & - & - & 0(6) &

0(1) & -12 & 6 & -36 & -60 & 18 & 36 & 15 & -9 & -3 & 12 & 20 & 5

& 0 & 1  \\  \hline

$c$ & 4 & ( 3,-1) & ( 0, 1) & ( 1,-1) & 2 & -2 & - & - & - & - &

-18 & -36 & -15 & 9 & 12 & -6 & 36 & 60 & -5 & -20 & -12 & 3 & -1

& 0  \\ \hline \hline

$d$ & 2 & ( 9,-1) & (-3,-1) & (-2, 0) & 12 & -12 & - & - & - & - &

- & - & 0 & 0 & -288 & 0 & 462 & 78 & -12 & -42 & -90 & -60 & 0 &

-6    \\ \hline

$e$ & 2 & ( 3, 4) & ( 1,-4) & ( 2, 0) & -16 & 16 & - & - & - & - &

- & - & - & - & 462 & -78 & -720 & 0 & 276 & -204 & -140 & 340 & 0

& 8  \\ \hline

$f$ & 2 & ( 9, 1) & (-1,-3) & ( 0, 2) & -12 & 12 & - & - & - & - &

- & - & - & - & - & - & 0 & 0 & 60 & 90 & 42 & 12 & 6 & 0  \\

\hline

$g$ & 2 & ( 3,-4) & (-4, 1) & ( 0,-2) & 16 & -16 & - & - & - & - &

- & - & - & - & - & - & - & - & -340 & 140 & 204 & -276 & -8 & 0

\\ \hline

$h$ & 2 & ( 1, 0) & ( 5, 3) & ( 5,-3) & 0 & 0 & - & - & - & - & -

& - & - & - & - & - & - & - & - & - & 0 & 0 & 0 & 0   \\ \hline

$i$ & 2 & ( 1, 0) & ( 3,-5) & ( 3, 5) & 0 & 0 & - & - & - & - & -

& - & - & - & - & - & - & - & - & - & - & - & 0 & 0  \\ \hline

$O6^{1}$ & 4 & ( 1, 0) & ( 1, 0) & ( 2, 0) & - & - & - & - & - & -

& - & - & - & - & - & - & - & - & - & - & - & - & - & - \\ \hline

$O6^{2}$ & 4 & ( 1, 0) & ( 0,-1) & ( 0, 2) & - & - & - & - & - & -

& - & - & - & - & - & - & - & - & - & - & - & - & - & - \\ \hline

\end{tabular}
\caption{D6-brane configurations and intersection numbers
for Model TI-U-7 on Type IIA $\mathbf{T}^6 / \Z_2 \times \Z_2$ orientifold.
The complete gauge symmetry is
$[U(4)_C \times U(2)_L \times U(2)_R]_{observable} \times
[U(1)^6 \times USp(4)^2]_{hidden}$, the
SM fermions and Higgs fields arise from
the intersections on the first two-torus,
and the complex structure parameters are
$2\chi_1=6\chi_2=3\chi_3 =6$.
To satisfy the RR tadpole cancellation conditions,
we choose $h_0=-6(q+8)$, $m=8$, and $a=16$.
A lot of exotic particles arise from extra gauge
groups due to the large supergravity fluxes
and the RR tadpole cancellation conditions. }
\label{TI-U-7}
\end{center}
\end{table}

\newpage


\subsection{$U(4)_C\times USp(2)_L \times U(2)_R$ Models}

\begin{table}[h]

\begin{center}

\footnotesize

\begin{tabular}{|@{}c@{}|c||@{}c@{}c@{}c@{}||c|c||c|c|c|@{}c@{}||
c|@{}c@{}|@{}c@{}|@{}c@{}|@{}c@{}|@{}c@{}|@{}c@{}|@{}c@{}|@{}c@{}|@{}c@{}|@{}c@{}|}

\hline

stack & $N$ & ($n_1$,$l_1$) & ($n_2$,$l_2$) & ($n_3$,$l_3$) & A & S

& $b$ & $b'$ & $c$ & $c'$ & $d$ & $d'$ & $e$ & $e'$ & $f$ & $f'$ &

$g$ & $g'$ & $h$ & $h'$ & $O6$     \\ \hline \hline

$a$ & 4 & ( 1, 0) & (-1,-1) & (-3, 1) & 0 & 0 & 3 & - & -3 & 0(1)

& -2 & 0(1) & 1 & -2 & 2 & 0(2) & 2 & 1 & 0(2) & 0(8) & 0(1)  \\

\hline

$b$ & 1 & ( 0,-1) & ( 1, 0) & ( 0, 2) & 0 & 0 & - & - & 3 & - & 6

& - & -3 & - & 0(1) & - & 0(1) & - & -3 & - & 0(2) \\

\hline

$c$ & 2 & (-1,-1) & ( 0, 1) & ( 3, 1) & 2 & -2 & - & - & - & - & 2

& 4 & 0(8) & 0(2) & -2 & -1 & -2 & 0(2) & 2 & -1 & 1  \\ \hline
\hline

$d$ & 2 & (-3,-1) & (-1, 1) & ( 2, 0) & 0 & 0 & - & - & - & - & -

& - & 6 & 0(1) & 0(3) & -6 & -4 & -2 & 4 & 2 & 0(1)  \\ \hline

$e$ & 1 & ( 3,-1) & ( 0, 1) & ( 1,-1) & -2 & 2 & - & - & - & - & -

& - & - & - & 0(3) & 3 & 2 & 0(4) & 0(1) & -1 & -1  \\ \hline

$f$ & 1 & ( 0,-1) & ( 1,-1) & (-1, 1) & -2 & 2 & - & - & - & - & -

& - & - & - & - & - & -1 & 0(1) & 0(4) & 2 & -1  \\ \hline

$g$ & 1 & ( 1,-1) & ( 1, 0) & ( 1, 1) & 0 & 0 & - & - & - & - & -

& - & - & - & - & - & - & - & -3 & 0(3) & 0(1)   \\ \hline

$h$ & 1 & ( 1, 0) & (-1,-3) & (-1, 1) & 0 & 0 & - & - & - & - & -

& - & - & - & - & - & - & - & - & - & 0(3)   \\ \hline

$O6$ & 1 & ( 1, 0) & ( 1, 0) & ( 2, 0) & - & - & - & - & - & - & -

& - & - & - & - & - & - & - & - & - & -   \\ \hline

\end{tabular}

\caption{D6-brane configurations and intersection numbers
for Model TI-Sp-1 on Type IIA $\mathbf{T^6}$ orientifold.
The complete gauge symmetry is
$[U(4)_C \times USp(2)_L \times U(2)_R]_{observable}
\times [U(2) \times U(1)^4 \times USp(2)]_{hidden}$, the
SM fermions and Higgs fields arise from
the intersections on the third two-torus,
and the complex structure parameters are
$2\chi_1= 6\chi_2= \chi_3=2\sqrt{3}$.
To satisfy the RR tadpole cancellation conditions,
we choose $h_0=-6(q+2)$, $m=2$, and $a=4$.}
\label{TI-Sp-1}
\end{center}
\end{table}

\newpage


\begin{table}[h]

\begin{center}

\footnotesize

\begin{tabular}{|@{}c@{}|c||@{}c@{}c@{}c@{}||c|c||c|c|c|@{}c@{}||
c|@{}c@{}|@{}c@{}|c|@{}c@{}|@{}c@{}|@{}c@{}|c|@{}c@{}|c|@{}c@{}|c|}

\hline

stack & $N$ & ($n_1$,$l_1$) & ($n_2$,$l_2$) & ($n_3$,$l_3$) & A & S

& $b$ & $b'$ & $c$ & $c'$ & $d$ & $d'$ & $e$ & $e'$ & $f$ & $f'$

& $g$ & $g'$ & $h$ & $h'$ & $i$ & $i'$ \\

\hline \hline

$a$ & 4 & ( 1, 1) & (-1,-1) & (-1, 3) & 12 & 0 & 3 & - & -3 & 0(9)

& 6 & 0(3) & -36 & -36 & 0(2) & 0(3) & 52 & 22 & 1 & - & -3 & -

\\ \hline

$b_{O6}$ & 1 & ( 1, 0) & ( 2, 0) & ( 1, 0) & - & - & - & - & 3 &

- & -6 & - & 0(25) & - & 0(1) & - & 0(1) & - & 0(1) & -

& 0(2) & - \\ \hline

$c$ & 2 & ( 0,-1) & ( 1,-1) & (-2, 3) & -6 & 6 & - & - & - & - &

18 & 6 & 6 & -9 & -3 & 0(3) & -42 & -30 & 0(2) & - & 0(3) & -

\\ \hline  \hline

$d$ & 1 & ( 2, 1) & ( 0, 2) & (-1,-3) & 12 & -12 & - & - & - & - &

- & - & -27 & 33 & 9 & -3 & 110 & 26 & -4 & - & 0(6) & - \\ \hline

$e$ & 1 & ( 1, 5) & ( 1,-5) & ( 1, 0) & 0 & 0 & - & - & - & - & -

& - & - & - & 0(18) & 0(8) & -80 & 70 & -5 & - & 0(1) & - \\

\hline

$f$ & 1 & ( 1,-1) & ( 1, 1) & ( 1, 0) & 0 & 0 & - & - & - & - & -

& - & - & - & - & - & -2 & 4 & 1 & - & 0(1) & -  \\ \hline

$g$ & 1 & ( 3,-1) & ( 2, 0) & ( 4, 1) & 0 & 0 & - & - & - & - &

- & - & - & - & - & - & - & - & 0(12) & - & 6 & -  \\ \hline

$h$ & 2 & ( 0,-1) & ( 2, 0) & ( 0, 1) & 0 & 0 & - & - & - & - & -

& - & - & - & - & - & - & - & - & - & 0(2) & -   \\ \hline

$i$ & 2 & ( 0,-1) & ( 0, 2) & ( 1, 0) & 0 & 0 & - & - & - & - & -

& - & - & - & - & - & - & - & - & - & - & -  \\ \hline

\end{tabular}

\caption{D6-brane configurations and intersection numbers
for Model TI-Sp-2 on Type IIA $\mathbf{T^6}$ orientifold.
The complete gauge symmetry is
$[U(4)_C \times USp(2)_L \times U(2)_R]_{observable}
\times [U(1)^4\times USp(4)^2]_{hidden}$, the
SM fermions and Higgs fields arise from
the intersections on the third two-torus,
and the complex structure parameters are
$4 \chi_1 = 2\chi_2 = 3\chi_3 =2\sqrt{2}$.
To satisfy the RR tadpole cancellation conditions,
we choose $h_0=-4(3q+4)$, $m=2$, and $a=8$.}
\label{TI-Sp-2}
\end{center}
\end{table}

\newpage


\begin{table}[h]

\begin{center}

\footnotesize

\begin{tabular}{|@{}c@{}|c||@{}c@{}c@{}c@{}||c|c||c|c|c|@{}c@{}||
@{}c@{}|@{}c@{}|@{}c@{}|@{}c@{}|c|@{}c@{}|@{}c@{}|c|} \hline

stack & $N$ & ($n_1$,$l_1$) & ($n_2$,$l_2$) & ($n_3$,$l_3$) & A &

S & $b$ & $b'$ & $c$ & $c'$ & $d$ & $d'$ & $e$ & $e'$ & $f$ & $f'$

& $g$ & $g'$   \\ \hline \hline

$a$ & 4 & ( 0,-1) & ( 1, 3) & ( 3, 1) & 3 & -3 & 3 & - & -3 & 0(2)

& 2 & 1 & 15 & 6 & -3 & 0(1) & 0(1) & - \\ \hline

$b_{O6}$ & 1 & ( 1, 0) & ( 2, 0) & ( 2, 0) & - & - & - & - & 3 &

- & 0(1) & - & 0(1) & - & 6 & - & 0(2) & -  \\ \hline

$c$ & 2 & ( 1,-1) & (-1, 3) & (-1,-1) & -6 & 0 & - & - & - & - &

-2 & 0(2) & -24 & 0(9) & 4 & 4 & 1 & -  \\ \hline  \hline

$d$ & 2 & ( 1, 1) & ( 1,-1) & ( 2, 0) & 0 & 0 & - & - & - & - & -

& - & 0(1) & -2 & 4 & -2 & 0(1) & -   \\ \hline

$e$ & 1 & ( 1, 1) & ( 2, 0) & ( 7,-1) & 0 & 0 & - & - & - & - & -

& - & - & - & 16 & -20 & -2 & -  \\ \hline

$f$ & 1 & ( 1,-3) & ( 0, 2) & ( 3,-1) & -6 & 6 & - & - & - & - & -

& - & - & - & - & - & 0(1) & -  \\ \hline

$g$ & 1 & ( 0,-1) & ( 0, 2) & ( 2, 0) & 0 & 0 & - & - & - & - & -

& - & - & - & - & - & - & -  \\ \hline

\end{tabular}
\caption{D6-brane configurations and intersection numbers
for Model TI-Sp-3 on Type IIA $\mathbf{T^6}$ orientifold.
The complete gauge symmetry is
$[U(4)_C \times USp(2)_L \times U(2)_R]_{observable}
\times [U(2) \times U(1)^2 \times USp(2)]_{hidden}$, the
SM fermions and Higgs fields arise from
the intersections on the second two-torus,
and the complex structure parameters are
$14\chi_1 = 7\chi_2 = \chi_3=2\sqrt{7}$.
To satisfy the RR tadpole cancellation conditions,
we choose $h_0=-4(3q+2)$, $m=2$, and $a=8$.}
\label{TI-Sp-3}
\end{center}
\end{table}

\newpage


\begin{table}[h]

\begin{center}

\footnotesize

\begin{tabular}{|@{}c@{}|c||@{}c@{}c@{}c@{}||c|c||c|c|@{}c@{}|@{}c@{}||
@{}c@{}|c|c|c|@{}c@{}|c|} \hline

stack & $N$ & ($n_1$,$l_1$) & ($n_2$,$l_2$) & ($n_3$,$l_3$) & A &

S & $b$ & $b'$ & $c$ & $c'$ & $d$ & $d'$ & $e$ & $e'$ & $f$ & $f'$

  \\ \hline \hline

$a$ & 4 & ( 0,-1) & ( 1, 3) & ( 3, 1) & 3 & -3 & 3 & - & -3 & 0(2)

& 12 & 15 & 3 & -6 & 0(1) & - \\ \hline

$b_{O6}$ & 1 & ( 1, 0) & ( 2, 0) & ( 2, 0) & - & - & - & - & 0(3)

& - & 0(2) & - & -12 & - & 0(2) & -  \\ \hline

$c$ & 2 & ( 1, 0) & ( 1,-3) & ( 1, 1) & 0 & 0 & - & - & - & - & 10

& -8 & -12 & -6 & 1 & -

\\ \hline  \hline

$d$ & 1 & ( 3,-2) & ( 3, 1) & ( 2, 0) & 0 & 0 & - & - & - & - & -

& - & -40 & 64 & 0(9) & -

\\ \hline

$e$ & 1 & ( 1, 6) & ( 1, 1) & ( 0,-2) & 12 & -12 & - & - & - & - &

- & - & - & - & -2 & -

\\ \hline

$f$ & 2 & ( 0,-1) & ( 0, 2) & ( 2, 0) & 0 & 0 & - & - & - & - & -

& - & - & - & - & -

\\ \hline

\end{tabular}

\caption{D6-brane configurations and intersection numbers
for Model TI-Sp-4 on Type IIA $\mathbf{T^6}$ orientifold.
The complete gauge symmetry is
$[U(4)_C \times USp(2)_L \times U(2)_R]_{observable}
\times [U(1)^2 \times USp(4)]_{hidden}$, the
SM fermions and Higgs fields arise from
the intersections on the second two-torus,
and the complex structure parameters are
$12\chi_1= 3\chi_2 = \chi_3 =2\sqrt{3}$.
To satisfy the RR tadpole cancellation conditions,
we choose $h_0=-4(3q+2)$, $m=2$, and $a=8$.}
\label{TI-Sp-4}
\end{center}
\end{table}

\newpage


\section{ The D6-Brane Configurations and Intersection Numbers
for  the Second Kind of Pati-Salam Models}
In Appendix B, we present the D6-brane configurations and
intersection numbers
for  the second kind of Pati-Salam models.

\subsection{$U(4)_C\times U(2)_L \times U(2)_R$ Models}

\begin{table}[h]

\begin{center}

\footnotesize

\begin{tabular}{|@{}c@{}|c||@{}c@{}c@{}c@{}||c|c||c|@{}c@{}|c|
@{}c@{}||@{}c@{}|@{}c@{}|@{}c@{}|@{}c@{}|@{}c@{}|@{}c@{}|@{}c@{}|@{}c@{}|@{}c@{}|c|}

\hline

stack & $N$ & ($n_1$,$l_1$) & ($n_2$,$l_2$) & ($n_3$,$l_3$) & A &

S & $b$ & $b'$ & $c$ & $c'$ & $d$ & $d'$ & $e$ & $e'$ & $f$ & $f'$

& $g$ & $g'$ & $h$ & $h'$

\\ \hline \hline

$a$ & 4 & ( 1, 0) & (-1,-3) & (-1, 3) & 0 & 0 & 3 & 0(1) & -3 &

0(1) & 0(1) & 3 & 6 & 3 & 0(3) & -9 & -6 & 3 & 0(1) & - \\ \hline

$b$ & 2 & ( 1,-1) & ( 1,-3) & ( 0, 2) & -6 & 6 & - & - & 2 & 0(1)

& 0(6) & 0(0) & 0(2) & 0(8) & -6 & 0(6) & -14 & -20 & 0(1) & - \\

\hline

$c$ & 2 & (-1,-1) & ( 0, 2) & ( 1, 3) & 6 & -6 & - & - & - & - &

0(1) & -2 & -4 & -2 & 9 & 3 & 15 & 21 & 0(1) & -

\\ \hline \hline

$d$ & 1 & ( 1, 1) & ( 1, 3) & ( 0,-2) & 6 & -6 & - & - & - & - & -

& - & 0(8) & 0(2) & 0(6) & 6 & 20 & 14 & 0(1) & - \\ \hline

$e$ & 1 & ( 1,-3) & (-1, 1) & ( 0,-2) & -6 & 6 & - & - & - & - & -

& - & - & - & -20 & 14 & 0(19) & -34 & 0(3) & -  \\ \hline

$f$ & 1 & ( 2,-1) & ( 1, 3) & ( 2, 0) & 0 & 0 & - & - & - & - & -

& - & - & - & - & - & 0(16) & 0(4) & -2 & -  \\ \hline

$g$ & 1 & ( 6, 1) & ( 1,-1) & ( 2, 0) & 0 & 0 & - & - & - & - & -

& - & - & - & - & - & - & - & 2 & -  \\ \hline

$h$ & 1 & ( 1, 0) & ( 0,-2) & ( 0, 2) & 0 & 0 & - & - & - & - & -

& - & - & - & - & - & - & - & - & -  \\ \hline

\end{tabular}
\caption{D6-brane configurations and intersection numbers
for Model TII-U-1 on Type IIA $\mathbf{T^6}$ orientifold.
The complete gauge symmetry is
$[U(4)_C \times U(2)_L \times U(2)_R]_{observable} \times
[U(1)^4 \times USp(2)]_{hidden}$,
the SM fermions and Higgs fields arise from
the intersections on different two-tori,
and the complex structure parameters are
$\chi_1 =3\chi_2=3\chi_3= \sqrt{2}$.
To satisfy the RR tadpole cancellation conditions,
we choose $h_0=-2(3q+2)$, $m=2$, and $a=4$.}
\label{TII-U-1}
\end{center}
\end{table}

\newpage



\begin{table}[h]

\begin{center}

\footnotesize

\begin{tabular}{|@{}c@{}|c||@{}c@{}c@{}c@{}||c|c||c|@{}c@{}|c|
@{}c@{}||@{}c@{}|@{}c@{}|@{}c@{}|@{}c@{}|@{}c@{}|@{}c@{}|@{}c@{}|@{}c@{}|@{}c@{}|}

\hline

stack & $N$ & ($n_1$,$l_1$) & ($n_2$,$l_2$) & ($n_3$,$l_3$) & A &

S & $b$ & $b'$ & $c$ & $c'$ & $d$ & $d'$ & $e$ & $e'$ & $f$ & $f'$

& $g$ & $g'$ & $O6$

\\ \hline \hline

$a$ & 4 & ( 2, 0) & ( 3, 1) & ( 3,-1) & 0 & 0 & 3 & 0(1) & -3 &

0(1) & 2 & 1 & 0(1) & 0(4) & 0(6) & 0(6) & -9 & 0(3) & 0(1) \\

\hline

$b$ & 2 & ( 3,-1) & ( 2, 0) & ( 3, 1) & 0 & 0 & - & - & 3 & 0(1)

& 0(1) & 0(4) & -2 & -1 & 9 & 0(3) & 0(6) & 0(6) & 0(1) \\

\hline

$c$ & 2 & ( 3, 1) & ( 3,-1) & ( 2, 0) & 0 & 0 & - & - & - & - & -2

& -1 & 2 & 1 & 0(3) & -9 & 0(3) & 9 & 0(1) \\ \hline \hline

$d$ & 3 & ( 1,-1) & ( 2, 0) & ( 1, 1) & 0 & 0 & - & - & - & - & -

& - & -1 & 0(1) & 2 & 1 & 4 & -4 & 0(1)  \\ \hline

$e$ & 1 & ( 2, 0) & ( 1, 1) & ( 1,-1) & 0 & 0 & - & - & - & - & -

& - & - & - & -4 & 4 & -2 & -1 & 0(1)   \\ \hline

$f$ & 1 & ( 0, 2) & ( 3,-1) & ( 3,-1) & -1 & 1 & - & - & - & - & -

& - & - & - & - & - & 27 & 0(9) & -2  \\ \hline

$g$ & 1 & ( 3, 1) & ( 0,-2) & ( 3, 1) & 1 & -1 & - & - & - & - & -

& - & - & - & - & - & - & - & 2  \\ \hline

$O6$ & 1 & ( 2, 0) & ( 2, 0) & ( 2, 0) & - & - & - & - & - & - & -

& - & - & - & - & - & - & - & -  \\ \hline

\end{tabular}

\caption{D6-brane configurations and intersection numbers
for Model TII-U-2 on Type IIA $\mathbf{T^6}$ orientifold.
The complete gauge symmetry is
$[U(4)_C \times U(2)_L \times U(2)_R]_{observable} \times
[U(3)\times U(1)^3 \times USp(2)]_{hidden}$,
the SM fermions and Higgs fields arise from
the intersections on different two-tori,
and the complex structure parameters are
$\chi_1 = \chi_2 = \chi_3= 6$.
To satisfy the RR tadpole cancellation conditions,
we choose $h_0=-36(q+4)$, $m=2$, and $a=24$.}
\label{TII-U-2}
\end{center}
\end{table}

\newpage


\begin{table}[h]

\begin{center}

\footnotesize

\begin{tabular}{|@{}c@{}|@{}c@{}||@{}c@{}c@{}c@{}||c|c||c|@{}c@{}|c|@{}c@{}||
@{}c@{}|@{}c@{}|@{}c@{}|@{}c@{}|@{}c@{}|@{}c@{}|@{}c@{}|@{}c@{}|@{}c@{}|c|@{}c@{}|c|@{}c@{}|}

\hline

stack & $N$ & ($n_1$,$l_1$) & ($n_2$,$l_2$) & ($n_3$,$l_3$) & A & S

& $b$ & $b'$ & $c$ & $c'$ & $d$ & $d'$ & $e$ & $e'$ & $f$ & $f'$

& $g$ & $g'$ & $h$ & $h'$ & $i$ & $i'$ & $O6$ \\

\hline \hline

$a$ & 4 & ( 1, 0) & (-1,-3) & (-1, 3) & 0 & 0 & 3 & 0(1) & -3 &

0(1) & 3 & 6 & 6 & 3 & 0(9) & -27 & -6 & 3 & 0(1) & - & 3 & - &

0(9)  \\ \hline

$b$ & 2 & ( 2,-1) & ( 1,-3) & ( 0, 2) & -6 & 6 & - & - & 4 & 0(1)

& 0(16) & 0(4) & 0(4) & 0(16) & 12 & 0(20) & -12 & -8 & 0(1) & - &

0(6) & - & -6   \\ \hline

$c$ & 2 & (-2,-1) & ( 0, 2) & ( 1, 3) & 6 & -6 & - & - & - & - &

-4 & -8 & -8 & -4 & 30 & -6 & 6 & 18 & 0(1) & - & -4 & - & 6

\\ \hline  \hline

$d$ & 1 & ( 2, 3) & ( 1, 1) & ( 0,-2) & 6 & -6 & - & - & - & - & -

& - & 0(12) & 0(0) & -36 & 24 & 20 & 0(28) & 0(3) & - & 0(2) & - &

6 \\ \hline

$e$ & 1 & ( 2,-3) & ( 1,-1) & ( 0, 2) & -6 & 6 & - & - & - & - & -

& - & - & - & -24 & 36 & 0(28) & -20 & 0(3) & - & 0(2) & - & -6

\\ \hline

$f$ & 1 & ( 4,-3) & ( 1, 3) & ( 2, 0) & 0 & 0 & - & - & - & - & -

& - & - & - & - & - & 0(32) & 0(8) & -6 & - & 24 & - & 0(9) \\

\hline

$g$ & 1 & ( 4, 1) & ( 1,-1) & ( 2, 0) & 0 & 0 & - & - & - & - & -

& - & - & - & - & - & - & - & 2 & - & -8 & - & 0(1)  \\ \hline

$h$ & 1 & ( 1, 0) & ( 0,-2) & ( 0, 2) & 0 & 0 & - & - & - & - & -

& - & - & - & - & - & - & - & - & - & 0(2) & - & 0(4) \\ \hline

$i$ & 2 & ( 0,-1) & ( 2, 0) & ( 0, 2) & 0 & 0 & - & - & - & - & -

& - & - & - & - & - & - & - & - & - & - & - & 0(2) \\ \hline

$O6$ & 3 & ( 1, 0) & ( 2, 0) & ( 2, 0) & - & - & - & - & - & - & -

& - & - & - & - & - & - & - & - & - & - & - & -

\\ \hline

\end{tabular}
\caption{D6-brane configurations and intersection numbers
for Model TII-U-3 on Type IIA $\mathbf{T^6}$ orientifold.
The complete gauge symmetry is
$[U(4)_C \times U(2)_L \times U(2)_R]_{observable} \times
[U(1)^4 \times USp(2) \times USp(4) \times USp(6)]_{hidden}$,
the SM fermions and Higgs fields arise from
the intersections on different two-tori,
and the complex structure parameters are
$\chi_1 = 2\chi_2 = 2\chi_3 =2\sqrt{6}/3$.
To satisfy the RR tadpole cancellation conditions,
we choose $h_0=-4(3q+4)$, $m=2$, and $a=8$.}
\label{TII-U-3}
\end{center}
\end{table}

\newpage


\begin{table}[h]

\begin{center}

\footnotesize

\begin{tabular}{|@{}c@{}|c||@{}c@{}c@{}c@{}||c|c||c|@{}c@{}|c|@{}c@{}||
@{}c@{}|@{}c@{}|@{}c@{}|@{}c@{}|@{}c@{}|@{}c@{}|@{}c@{}|@{}c@{}|@{}c@{}|c|}

\hline

stack & $N$ & ($n_1$,$l_1$) & ($n_2$,$l_2$) & ($n_3$,$l_3$) & A & S

& $b$ & $b'$ & $c$ & $c'$ &

$d$ & $d'$ & $e$ & $e'$ & $f$ & $f'$ & $g$ & $g'$ & $h$ & $h'$  \\

\hline \hline

$a$ & 4 & ( 1, 0) & (-1,-1) & (-1, 3) & 0 & 0 & 3 & 0(3) & -3 &

0(2) & -1 & 2 & 0(1) & 3 & 0(3) & -9 & 2 & -1 & 0(1) & -

\\ \hline

$b$ & 2 & ( 2,-3) & ( 1,-1) & ( 0, 2) & -6 & 6 & - & - & 4 &

2 & 0(16) & 0(4) & -8 & -4 & 0(1) & 12 & -16 & -20 & 0(3) & - \\

\hline

$c$ & 2 & ( 0,-1) & ( 1, 3) & ( 1, 3) & 9 & -9 & - & - & - & - &

0(2) & -6 & 18 & 0(6) & -6 & 0(2) & 18 & 36 & -1 & - \\ \hline \hline

$d$ & 2 & ( 2, 1) & ( 1, 3) & ( 0,-2) & 6 & -6 & - & - & - & - & -

& - & 0(3) & 12 & -8 & -4 & 24 & 12 & 0(1) & -  \\ \hline

$e$ & 2 & (-2,-1) & ( 2, 0) & (-1, 3) & 0 & 0 & - & - & - & - & -

& - & - & - & 0(16) & -24 & 0(16) & 0(4) & 2 & -  \\ \hline

$f$ & 1 & ( 2,-3) & ( 0,-2) & (-1, 3) & -18 & 18 & - & - & - & - &

- & - & - & - & - & - & -64 & -40 & 0(3) & -  \\ \hline

$g$ & 1 & ( 6,-1) & ( 2, 0) & ( 1, 1) & 0 & 0 & - & - & - & - & -

& - & - & - & - & - & - & - & -2 & -  \\ \hline

$h$ & 4 & ( 1, 0) & ( 0,-2) & ( 0, 2) & 0 & 0 & - & - & - & - & -

& - & - & - & - & - & - & - & - & -  \\ \hline

\end{tabular}
\caption{D6-brane configurations and intersection numbers
for Model TII-U-4 on Type IIA $\mathbf{T^6}$ orientifold.
The complete gauge symmetry is
$[U(4)_C \times U(2)_L \times U(2)_R]_{observable} \times
[U(2)^2 \times U(1)^2 \times USp(8)]_{hidden}$,
the SM fermions and Higgs fields arise from
the intersections on different two-tori,
and the complex structure parameters are
$\chi_1 = \chi_2 = 3\chi_3 = 2/\sqrt{3}$.
To satisfy the RR tadpole cancellation conditions,
we choose $h_0=-4(3q+2)$, $m=2$, and $a=8$.}
\label{TII-U-4}
\end{center}
\end{table}

\newpage



\begin{table}[h]

\begin{center}

\footnotesize

\begin{tabular}{|@{}c@{}|c||@{}c@{}c@{}c@{}||c|c||c|@{}c@{}|@{}c@{}|@{}c@{}||
c|c|c|@{}c@{}|@{}c@{}|c|@{}c@{}|c|} \hline

stack & $N$ & ($n_1$,$l_1$) & ($n_2$,$l_2$) & ($n_3$,$l_3$) & A & S

& $b$ & $b'$ & $c$ & $c'$ &

$d$ & $d'$ & $e$ & $e'$ & $f$ & $f'$ & $g$ & $g'$   \\

\hline \hline

$a$ & 4 & ( 1, 0) & (-1,-3) & (-1, 1) & 0 & 0 & 3 & 0(1) & -3 &

0(1) & -12 & -6 & -6 & 12 & 0(1) & - & 3 & -

\\ \hline

$b$ & 2 & ( 1,-1) & ( 1,-3) & ( 0, 1) & -6 & 6 & - & - & 0(2) &

0(3) & 12 & 6 & -36 & -18 & 0(1) & - & 0(3) & - \\

\hline

$c$ & 2 & ( 1, 1) & (-1, 3) & (-1, 0) & 0 & 0 & - & - & - & - & -2

& 4 & 6 & -12 & 1 & - & -3 & -  \\ \hline  \hline

$d$ & 1 & ( 1, 3) & ( 0,-2) & ( 3, 1) & 12 & -12 & - & - & - & - &

- & - &  96 & 0(20) & 0(9) & - & -6 & -

\\ \hline

$e$ & 1 & ( 3, 1) & ( 2, 0) & ( 3,-1) & 0 & 0 & - & - & - & - & -

& - & - & - & 6 & - & 0(9) & -

\\ \hline

$f$ & 2 & ( 1, 0) & ( 0,-2) & ( 0, 1) & 0 & 0 & - & - & - & - & -

& - & - & - & - & - & 0(2) & -

\\ \hline

$g$ & 2 & ( 0,-1) & ( 2, 0) & ( 0, 1) & 0 & 0 & - & - & - & - & -

& - & - & - & - & - & - & -

\\ \hline

\end{tabular}
\caption{D6-brane configurations and intersection numbers
for Model TII-U-5 on Type IIA $\mathbf{T^6}$ orientifold.
The complete gauge symmetry is
$[U(4)_C \times U(2)_L \times U(2)_R]_{observable} \times
[U(1)^2 \times USp(4)^2]_{hidden}$,
the SM fermions and Higgs fields arise from
the intersections on different two-tori,
and the complex structure parameters are
$2\chi_1 = 3\chi_2 = 2\chi_3 = 2$.
To satisfy the RR tadpole cancellation conditions,
we choose $h_0 = -2(3q+4)$, $m=2$, and $a=4$.}
\label{TII-U-5}
\end{center}
\end{table}

\newpage



\begin{table}[h]

\begin{center}

\footnotesize

\begin{tabular}{|@{}c@{}|c||@{}c@{}c@{}c@{}||c|c||c|@{}c@{}|@{}c@{}|
@{}c@{}||@{}c@{}|c|@{}c@{}|c|@{}c@{}|c|} \hline

stack & $N$ & ($n_1$,$l_1$) & ($n_2$,$l_2$) & ($n_3$,$l_3$) & A & S

& $b$ & $b'$ & $c$ & $c'$ & $d$ & $d'$ & $e$ & $e'$ & $f$ & $f'$

\\ \hline \hline

$a$ & 4 & ( 1, 0) & ( 3, 1) & ( 3,-1) & 0 & 0 & 3 & 0(2) & -3 &

0(2) & -2 & -4 & 0(9) & - & 3 & -  \\ \hline

$b$ & 2 & ( 0, 1) & ( 3,-1) & ( 1,-1) & -1 & 1 & - & - & 0(4) &

0(1) & -2 & 1 & 3 & - & 0(1) & -  \\  \hline

$c$ & 2 & ( 0,-1) & ( 1, 1) & ( 3, 1) & 1 & -1 & - & - & - & - &

0(1) & 1 & -3 & - & 0(3) & -  \\ \hline \hline

$d$ & 2 & ( 1,-2) & ( 1, 1) & ( 2, 0) & 0 & 0 & - & - & - & - & -

& - & -4 & - & 2 & - \\ \hline

$e$ & 1 & ( 1, 0) & ( 0,-2) & ( 0, 2) & 0 & 0 & - & - & - & - & -

& - & - & - & 0(2) & -  \\ \hline

$f$ & 2 & ( 0,-1) & ( 2, 0) & ( 0, 2) & 0 & 0 & - & - & - & - & -

& - & - & - & - & -  \\ \hline

\end{tabular}

\caption{D6-brane configurations and intersection numbers
for Model TII-U-6 on Type IIA $\mathbf{T^6}$ orientifold.
The complete gauge symmetry is
$[U(4)_C \times U(2)_L \times U(2)_R]_{observable} \times
[U(2) \times USp(2)\times USp(4)]_{hidden}$,
the SM fermions and Higgs fields arise from
the intersections on different two-tori,
and the complex structure parameters are
$4\chi_1= \chi_2=\chi_3= 2\sqrt{3}$.
To satisfy the RR tadpole cancellation conditions,
we choose $h_0=-12(q+2)$, $m=2$, and $a=8$.}
\label{TII-U-6}
\end{center}
\end{table}

\newpage


\subsection{$U(4)_C\times USp(2)_L \times U(2)_R$ Models}

\begin{table}[h]

\begin{center}

\footnotesize

\begin{tabular}{|@{}c@{}|c||@{}c@{}c@{}c@{}||c|c||c|c|c|
@{}c@{}||@{}c@{}|c|@{}c@{}|@{}c@{}|@{}c@{}|@{}c@{}|@{}c@{}|
c|@{}c@{}|@{}c@{}|@{}c@{}|c|@{}c@{}|c|} \hline

stack & $N$ & ($n_1$,$l_1$) & ($n_2$,$l_2$) & ($n_3$,$l_3$) & A & S

& $b$ & $b'$ & $c$ & $c'$ & $d$ & $d'$ & $e$ & $e'$ & $f$ & $f'$ &

$g$ & $g'$ & $h$ & $h'$ & $i$ & $i'$ & $j$ & $j'$

\\ \hline \hline

$a$ & 4 & ( 0,-1) & ( 1, 3) & ( 3, 1) & 6 & -6 & 3 & - & -3 & 0(1)

& -6 & -9 & 3 & 0(5) & 0(2) & 0(8) & 42 & 30 & 12 & 6 & -3 & - &

0(9) & -  \\ \hline

$b_{O6}$ & 1 & ( 1, 0) & ( 1, 0) & ( 2, 0) & - & - & - & - & 1 &

- & 0(6) & - & 0(2) & - & -1 & - & 0(1) & - & 0(1) &

- & 0(2) & - & 0(2) & -  \\  \hline

$c$ & 2 & ( 1,-1) & ( 0, 1) & ( 3,-1) & -2 & 2 & - & - & - & - & 1

& -2 & 0(1) & -3 & 2 & 1 & -14 & -10 & -4 & -2 & 0(3) & - & 3 & -

\\ \hline \hline

$d$ & 2 & ( 1, 0) & ( 1, 6) & ( 1,-1) & 0 & 0 & - & - & - & - & -

& - & 0(4) & 0(16) & 5 & 0(7) & -13 & 11 & -7 & 5 & 0(1) & - & 6 &

-  \\ \hline

$e$ & 2 & ( 1, 0) & ( 1, 2) & ( 3,-1) & 0 & 0 & - & - & - & - & -

& - & - & - & 2 & 3 & -5 & 3 & -3 & 1 & 0(3) & - & 6 & -

\\ \hline

$f$ & 2 & ( 0,-1) & ( 1, 1) & ( 1, 1) & 2 & -2 & - & - & - & - & -

& - & - & - & - & - & 18 & 6 & 6 & 0(3) & -1 & - & 0(1) & -

\\ \hline

$g$ & 1 & ( 6, 1) & ( 2,-1) & ( 2, 0) & 0 & 0 & - & - & - & - & -

& - & - & - & - & - & - & - & 0(3) & 0(27) & 4 & - & -12 & -

\\ \hline

$h$ & 1 & ( 3, 1) & ( 1,-1) & ( 2, 0) & 0 & 0 & - & - & - & - & -

& - & - & - & - & - & - & - & - & - & 2 & - & -6 & - \\ \hline

$i$ & 3 & ( 1, 0) & ( 0,-1) & ( 0, 2) & 0 & 0 & - & - & - & - & -

& - & - & - & - & - & - & - & - & - & - & - & - & - \\ \hline

$j$ & 1 & ( 0, -1) & ( 1, 0) & ( 0, 2) & 0 & 0 & - & - & - & - & -

& - & - & - & - & - & - & - & - & - & - & - & - & -  \\ \hline

\end{tabular}

\caption{D6-brane configurations and intersection numbers
for Model TII-Sp-1 on Type IIA $\mathbf{T^6}$ orientifold.
The complete gauge symmetry is
$[U(4)_C \times USp(2)_L \times U(2)_R]_{observable}
\times [U(2)^3 \times U(1)^2 \times USp(6) \times USp(2)
]_{hidden}$, the SM fermions and Higgs fields arise from
the intersections on different two-tori,
and the complex structure parameters are
$\chi_1 = 3\chi_2 =\chi_3/4 = \sqrt{3/2}$.
To satisfy the RR tadpole cancellation conditions,
we choose $h_0=-6(q+4)$, $m=2$, and $a=4$.}
\label{TII-Sp-1}
\end{center}
\end{table}

\newpage


\begin{table}[h]

\begin{center}

\footnotesize

\begin{tabular}{|@{}c@{}|c||@{}c@{}c@{}c@{}||c|c||c|c|c|
@{}c@{}||@{}c@{}|@{}c@{}|@{}c@{}|@{}c@{}|@{}c@{}|@{}c@{}|@{}c@{}|c|@{}c@{}|}

\hline

stack & $N$ & ($n_1$,$l_1$) & ($n_2$,$l_2$) & ($n_3$,$l_3$) & A & S

& $b$ & $b'$ & $c$ & $c'$ & $d$ & $d'$ & $e$ & $e'$ & $f$ & $f'$ &

$g$ & $g'$ & $O6$      \\ \hline \hline

$a$ & 4 & ( 1, 0) & ( 3, 1) & ( 3,-1) & 0 & 0 & 3 & - & -3 & 0(1)

& 0(2) & 0(8) & 0(3) & 9 & 3 & 0(5) & -3 & - & 0(1) \\ \hline

$b$ & 1 & ( 0,-1) & ( 2, 0) & ( 0, 1) & 0 & 0 & - & - & 3 & - &

-1 & - & 0(2) & - & 0(2) & - & 0(2) & - & 0(1) \\  \hline

$c$ & 2 & ( 3, 1) & ( 3,-1) & ( 1, 0) & 0 & 0 & - & - & - & - & 2

& 1 & -3 & 0(5) & 0(3) & -9 & 0(9) & - & 0(1)  \\ \hline \hline

$d$ & 2 & ( 1, 0) & ( 1, 1) & ( 1,-1) & 0 & 0 & - & - & - & - & -

& - & 1 & 2 & -2 & 3 & -1 & - & 0(1)  \\ \hline

$e$ & 1 & ( 2, 1) & ( 3, 1) & ( 0,-1) & 2 & -2 & - & - & - & - & -

& - & - & - & -12 & 0(4) & -6 & - & 1  \\ \hline

$f$ & 1 & ( 0, 1) & ( 3,-1) & ( 2,-1) & -2 & 2 & - & - & - & - & -

& - & - & - & - & - & 0(3) & - & -1  \\ \hline

$g$ & 2 & ( 0,-1) & ( 0, 2) & ( 1, 0) & 0 & 0 & - & - & - & - & -

& - & - & - & - & - & - & - & 0(2)  \\ \hline

$O6$ & 4 & ( 1, 0) & ( 2, 0) & ( 1, 0) & - & - & - & - & - & - & -

& - & - & - & - & - & - & - & -  \\ \hline

\end{tabular}

\caption{D6-brane configurations and intersection numbers
for Model TII-Sp-2 on Type IIA $\mathbf{T^6}$ orientifold.
The complete gauge symmetry is
$[U(4)_C \times USp(2)_L \times U(2)_R]_{observable}
\times [U(2)\times U(1)^2 \times USp(4) \times USp(8)]_{hidden}$,
the SM fermions and Higgs fields arise from
the intersections on different two-tori,
and the complex structure parameters are
$2\chi_1 = \chi_2=2\chi_3=2\sqrt{6}$.
To satisfy the RR tadpole cancellation conditions,
we choose $h_0=-12(3q+4)$, $m=2$, and $a=24$.}
\label{TII-Sp-2}
\end{center}
\end{table}

\newpage


\begin{table}[h]

\begin{center}

\footnotesize

\begin{tabular}{|@{}c@{}|c||@{}c@{}c@{}c@{}||c|c||c|c|c|
@{}c@{}||@{}c@{}|@{}c@{}|@{}c@{}|@{}c@{}|@{}c@{}|@{}c@{}|@{}c@{}|
@{}c@{}|@{}c@{}|@{}c@{}|@{}c@{}|@{}c@{}|@{}c@{}|c|} \hline

stack & $N$ & ($n_1$,$l_1$) & ($n_2$,$l_2$) & ($n_3$,$l_3$) & A & S

& $b$ & $b'$ & $c$ & $c'$ & $d$ & $d'$ & $e$ & $e'$ & $f$ & $f'$ &

$g$ & $g'$ & $h$ & $h'$ & $i$ & $i'$ & $j$ & $j'$

\\ \hline \hline

$a$ & 4 & ( 0,-1) & ( 1, 3) & ( 3, 1) & 6 & -6 & 3 & - & -3 & 0(1)

& -2 & -1 & -12 & -15 & 33 & 21 & 5 & 1 & 0(3) & 9 & 9 & 0(3) &

0(9) & -  \\ \hline

$b_{O6}$ & 1 & ( 1, 0) & ( 1, 0) & ( 2, 0) & - & - & - & - & 3 &

- & 1 & - & -3 & - & 0(4) & - & 0(4) & - & 0(2) & - &

0(2) & - & 0(2) & -  \\  \hline

$c$ & 2 & ( 1,-3) & ( 0, 1) & ( 3,-1) & -6 & 6 & - & - & - & - &

0(2) & 0(8) & 0(50) & 0(32) & -33 & -21 & -5 & -1 & -3 & 0(5) &

0(5) & -3 & 3 & -   \\ \hline \hline

$d$ & 1 & ( 1,-1) & ( 0, 1) & ( 1,-1) & -2 & 2 & - & - & - & - & -

& - & 0(8) & 0(2) & -15 & -3 & -3 & 1 & 2 & -3 & -3 & 2 & 1 & -

\\ \hline

$e$ & 1 & ( 3, 1) & ( 0, 1) & (-1,-3) & 6 & -6 & - & - & - & - & -

& - & - & - & -27 & 81 & -15 & 21 & 28 & -25 & -25 & 28 & -3 & -

\\ \hline

$f$ & 1 & ( 3, 2) & ( 3,-2) & ( 2, 0) & 0 & 0 & - & - & - & - & -

& - & - & - & - & - & 0(16) & 0(64) & -16 & -8 & -8 & -16 & -12 &

- \\ \hline

$g$ & 1 & ( 1, 2) & ( 1,-2) & ( 2, 0) & 0 & 0 & - & - & - & - & -

& - & - & - & - & - & - & - & -8 & 0(2) & 0(2) & -8 & -4 & -

\\ \hline

$h$ & 1 & ( 1,-2) & (-1, 0) & (-3,-1) & 0 & 0 & - & - & - & - & -

& - & - & - & - & - & - & - & - & - & 0(12) & 0(0) & 0(3) & -

\\ \hline

$i$ & 1 & ( 1, 2) & ( 1, 0) & ( 3,-1) & 0 & 0 & - & - & - & - & -

& - & - & - & - & - & - & - & - & - & - & - & 0(3) & -   \\ \hline

$j$ & 2 & ( 0,-1) & ( 1, 0) & ( 0, 2) & 0 & 0 & - & - & - & - & -

& - & - & - & - & - & - & - & - & - & - & - & - & -  \\ \hline

\end{tabular}

\caption{D6-brane configurations and intersection numbers
for Model TII-Sp-3 on Type IIA $\mathbf{T^6}$ orientifold.
The complete gauge symmetry is
$[U(4)_C \times USp(2)_L \times U(2)_R]_{observable}
\times [U(1)^6 \times USp(4)]_{hidden}$,
the SM fermions and Higgs fields arise from
the intersections on different two-tori,
and the complex structure parameters are
$12\chi_1 = 12\chi_2= \chi_3= 2\sqrt{6}$.
To satisfy the RR tadpole cancellation conditions,
we choose $h_0=-6(q+2)$, $m=2$, and $a=4$.}
\label{TII-Sp-3}
\end{center}
\end{table}

\newpage


\begin{table}[h]

\begin{center}

\footnotesize

\begin{tabular}{|@{}c@{}|c||@{}c@{}c@{}c@{}||c|c||c|c|@{}c@{}|
@{}c@{}||@{}c@{}|c|c|@{}c@{}|@{}c@{}|@{}c@{}|@{}c@{}|@{}c@{}|@{}c@{}|c|}
\hline

stack & $N$ & ($n_1$,$l_1$) & ($n_2$,$l_2$) & ($n_3$,$l_3$) & A & S

& $b$ & $b'$ & $c$ & $c'$ & $d$ & $d'$ & $e$ & $e'$ & $f$ & $f'$ &

$g$ & $g'$ & $h$ & $h'$    \\ \hline \hline

$a$ & 4 & ( 0, 1) & (-1,-3) & ( 3, 1) & 6 & -6 & 3 & - & -3 & 0(7)

& 5 & 1 & -18 & 0(3) & 16 & -4 & 28 & 20 & -3 & -  \\ \hline

$b_{O6}$ & 1 & ( 1, 0) & ( 1, 0) & ( 2, 0) & - & - & - & - & 0(4)

& - & 0(2) & - & 12 & - & 0(25) & - & 0(1) & - &

0(2) & -  \\  \hline

$c$ & 2 & (-1, 0) & ( 1, 4) & (-3, 1) & 0 & 0 & - & - & - & - & -6

& 2 & 42 & 6 & -45 & -5 & -9 & 7 & 0(3) & -  \\ \hline \hline

$d$ & 2 & ( 1, 1) & ( 1,-2) & ( 2, 0) & 0 & 0 & - & - & - & - & -

& - & 6 & -10 & 0(9) & 0(49) & 0(9) & 0(25) & 2 & -  \\ \hline

$e$ & 1 & (-1, 2) & (-1, 3) & ( 0, 2) & -24 & 24 & - & - & - & - &

- & - & - & - & 36 & 16 & -90 & -98 & 0(2) & -  \\ \hline

$f$ & 1 & ( 2, 5) & (-1, 5) & (-2, 0) & 0 & 0 & - & - & - & - & -

& - & - & - & - & - & 0(162) & 0(242) & 10 & -  \\ \hline

$g$ & 1 & ( 4, 1) & ( 2,-1) & ( 2, 0) & 0 & 0 & - & - & - & - & -

& - & - & - & - & - & - & - & 4 & -  \\ \hline

$h$ & 1 & ( 1, 0) & ( 0,-1) & ( 0, 2) & 0 & 0 & - & - & - & - & -

& - & - & - & - & - & - & - & - & -  \\ \hline

\end{tabular}
\caption{D6-brane configurations and intersection numbers
for Model TII-Sp-4 on Type IIA $\mathbf{T^6}$ orientifold.
The complete gauge symmetry is
$[U(4)_C \times USp(2)_L \times U(2)_R]_{observable}
\times [U(2) \times U(1)^3 \times USp(2)]_{hidden}$,
 the SM fermions and Higgs fields arise from
the intersections on different two-tori,
and the complex structure parameters are
$12\chi_1 = 24\chi_2 = \chi_3 =4\sqrt{3}$.
To satisfy the RR tadpole cancellation conditions,
we choose $h_0=-8(3q+2)$, $m=2$, and $a=16$.}
\label{TII-Sp-4}
\end{center}
\end{table}

\newpage



\begin{table}[h]

\begin{center}

\footnotesize

\begin{tabular}{|@{}c@{}|c||@{}c@{}c@{}c@{}||c|c||c|c|c|
c||c@{}|c@{}|@{}c@{}|@{}c@{}|@{}c@{}|@{}c@{}|@{}c@{}|
@{}c@{}|@{}c@{}|@{}c@{}|@{}c@{}|@{}c@{}|c@{}|@{}c@{}|@{}c@{}|@{}c@{}|@{}c@{}|@{}c@{}|}
\hline

stack & $N$ & ($n_1$,$l_1$) & ($n_2$,$l_2$) & ($n_3$,$l_3$) & A & S

& $b$ & $b'$ & $c$ & $c'$ & $d$ & $d'$ & $e$ & $e'$ & $f$ & $f'$ &

$g$ & $g'$ & $h$ & $h'$ & $i$ & $i'$ & $j$ & $j'$ & $k$ & $k'$ &

$O6^{1}$ & $O6^{4}$

\\ \hline \hline

$a$ & 8 & ( 1, 0) & ( 3, 1) & ( 3,-1) & 0 & 0 & 3 & - & -3 & 0 &

-1 & -2 & -15 & -12 & 33 & 21 & 30 & 24 & 15 & -21 & 21 & -33 & 6

& -2 & -36 & -15 & 0 & -3

\\ \hline

$b_{O6^{3}}$ & 2 & ( 0,-1) & ( 2, 0) & ( 0, 1) & - & - & - & - & 3

& - & -1 & - & 3 & - & 0 & - & 0 & - & -8 & - & -12 & - & 0 & -

& 0 & - & 0 & 0  \\  \hline

$c$ & 4 & ( 3, 1) & ( 3,-1) & ( 1, 0) & 0 & 0 & - & - & - & - & 0

& 0 & 0 & 0 & -33 & -21 & -30 & -24 & 42 & 6 & 54 & 0 & 5 & -2 &

-15 & 12 & 0 & 0  \\ \hline \hline

$d$ & 2 & ( 1,-1) & (-1,-1) & (-1, 0) & 0 & 0 & - & - & - & - & -

& - & 0 & 0 & 3 & 15 & 6 & 12 & -6 & -10 & -6 & -12 & -1 & 0 & -1

& 2 & 0 & 0    \\ \hline

$e$ & 2 & ( 1, 3) & ( 1,-3) & ( 1, 0) & 0 & 0 & - & - & - & - & -

& - & - & - & -81 & 27 & -54 & 0 & 26 & 22 & 30 & 24 & -7 & -16 &

-3 & 0 & 0 & 0  \\ \hline

$f$ & 2 & ( 2,-3) & (-2, 0) & (-2,-3) & 0 & 0 & - & - & - & - & -

& - & - & - & - & - & 0 & 0 & -140 & -28 & -168 & 0 & -8 & -16 &

96 & 120 & 0 &  12   \\ \hline

$g$ & 2 & ( 1,-3) & (-2, 0) & (-1,-3) & 0 & 0 & - & - & - & - & -

& - & - & - & - & - & - & - & -110 & 26 & -144 & 60 & -5 & -7 & 51

& 57 & 0 & 6   \\ \hline

$h$ & 2 & ( 4,-1) & ( 0, 2) & ( 1,-2) & 4 & -4 & - & - & - & - & -

& - & - & - & - & - & - & - & - & - & 0 & 0 & 2 & 6 & 52 & -44 &

-4 & 0   \\ \hline

$i$ & 2 & ( 3,-1) & ( 0, 2) & ( 2,-3) & 6 & -6 & - & - & - & - & -

& - & - & - & - & - & - & - & - & - & - & - & 2 & 10 & 60 & -48 &

-6 & 0  \\ \hline

$j$ & 2 & ( 2,-1) & ( 1,-1) & ( 0, 1) & 1 & -1 & - & - & - & - & -

& - & - & - & - & - & - & - & - & - & - & - & - & - & 24 & 12 & -1

& 2  \\ \hline

$k$ & 2 & ( 0, 1) & (-1,-3) & ( 6, 1) & -3 & 3 & - & - & - & - & -

& - & - & - & - & - & - & - & - & - & - & - & - & - & - & - & 3 &

0

\\ \hline

$O6^{1}$ & 6 & ( 1, 0) & ( 2, 0) & ( 1, 0) & - & - & - & - & - & -

& - & - & - & - & - & - & - & - & - & - & - & - & - & - & - & - &

- & -   \\ \hline

$O6^{4}$ & 4 & ( 0,-1) & ( 0, 2) & ( 1, 0) & - & - & - & - & - & -

& - & - & - & - & - & - & - & - & - & - & - & - & - & - & - & - &

- & -   \\ \hline

\end{tabular}
\caption{D6-brane configurations and intersection numbers
for Model TII-Sp-5 on Type IIA
$\mathbf{T}^6 / \Z_2 \times \Z_2$ orientifold.
The complete gauge symmetry is
$[U(4)_C \times USp(2)_L \times U(2)_R]_{observable}
\times [U(1)^8 \times USp(6)\times USp(4)]_{hidden}$,
 the SM fermions and Higgs fields arise from
the intersections on different two-tori,
and the complex structure parameters are
$\chi_1=\chi_2/2=\chi_3=\sqrt{2}$.
To satisfy the RR tadpole cancellation conditions,
we choose $h_0=-4(3q+8)$, $m=8$, and $a=32$.
A lot of exotic particles arise from extra gauge
groups due to the large supergravity fluxes
and the RR tadpole cancellation conditions.}
\label{TII-Sp-5}
\end{center}
\end{table}

\end{document}